\documentclass[conference]{IEEEtran}
\usepackage{cite}
\usepackage{amsmath,amssymb,amsfonts}
\usepackage[titlenumbered, ruled, linesnumbered, vlined]{algorithm2e}
\usepackage{algorithmic}
\usepackage{booktabs}
\usepackage{graphicx}
\usepackage{textcomp}
\usepackage{xcolor}
\usepackage{listings}
\usepackage{subfigure}
\usepackage{url}
\usepackage{tabularx}
\usepackage{caption}
\usepackage{balance}
\usepackage[bookmarks=false]{hyperref}
\usepackage{multirow}

\def\fixme#1{\typeout{FIXED in page \thepage : {#1}}
\bgroup \color{red}{[FIXME: {#1}]} \egroup}

\def\BibTeX{{\rm B\kern-.05em{\sc i\kern-.025em b}\kern-.08em
    T\kern-.1667em\lower.7ex\hbox{E}\kern-.125emX}}
\let\oldnl\nl
\newcommand{\nonl}{\renewcommand{\nl}{\let\nl\oldnl}}
\begin{document}

\title{RT-Gang: Real-Time Gang Scheduling Framework for Safety-Critical Systems}

 \author{Waqar Ali, Heechul Yun\\
   University of Kansas, USA. \\
   \{wali, heechul.yun\}@ku.edu \\
 }

\maketitle

\begin{abstract}
  In this paper, we present RT-Gang: a novel real-time gang scheduling
  framework that enforces a \emph{one-gang-at-a-time} policy.
  We find that, in a multicore platform, co-scheduling 
  multiple parallel real-time tasks would require highly pessimistic
  worst-case execution time (WCET) and schedulability analysis---even when
  there are enough cores---due to contention in shared hardware
  resources such as cache and DRAM controller.
            
  In RT-Gang, all threads of a parallel real-time task form a
  real-time gang  and the scheduler globally enforces the
  one-gang-at-a-time scheduling policy to guarantee tight and accurate
  task WCET. To minimize under-utilization, we integrate a
  state-of-the-art memory bandwidth throttling framework to allow
  safe execution of best-effort tasks. Specifically, any idle cores,
  if exist, are used to schedule best-effort tasks but 
  their maximum memory bandwidth usages are strictly throttled to
  tightly bound interference to real-time gang tasks.

  We implement RT-Gang in the
  Linux kernel and evaluate it on two representative embedded
  multicore platforms using both synthetic and real-world DNN
  workloads. The results show that RT-Gang dramatically 
  improves system predictability and the overhead is negligible.

\end{abstract}


\section{Introduction}
Safety-critical embedded real-time systems in automotive and aviation
industries increasingly demand efficient, high-performance computing platforms
to execute compute and data intensive workloads (e.g., deep neural networks) in
real-time, while meeting size, weight, power and cost constraints
~\cite{robert2015keynote}.  However, engineers and researchers developing and
studying these safety-critical systems have had troubles to deal with modern
high-performance computer architectures because of their unpredictable and
extremely poor worst-case timing behaviors that are too complex to understand
and analyze~\cite{axer2014building}.

In a safety-critical real-time system, the use of high-performance multicore
platforms is challenging because shared hardware resources, such as cache and
memory controllers, can cause extremely high timing
variations~\cite{yun2015ospert,valsan2016taming}.  The timing unpredictability
is a serious problem in both automotive and aviation industries. For example,
Bosch, a major automotive supplier, recently announced ``predictability on
high-performance platforms'' as a major industrial challenge for which the
industry is actively seeking solutions from the research
community~\cite{bosch2019challenge}. In aviation, the problem was dubbed as
``one-out-of-m'' problem~\cite{kim2017attacking} because the current industry
practice is to disable all but one core as recommended by the Federal Aviation
Administration (FAA) for certification, which requires \emph{evidence of
bounded interference}~\cite{faa2016certification}.

Prior efforts to address the timing predictability problem have been largely
based on the two basic ideas: (1) designing simpler time-predictable
architectures and (2) partitioning shared resources. Unfortunately, simpler
architectures tend to trade-off too much performance in favor of
predictability, which we can no longer afford. Partitioning shared resources
improves predictability but cannot guarantee tight worst-case timing in
high-performance multicore architectures because there are too many important
but unpartitionable shared resources~\cite{valsan2016taming}. Moreover, not
only partitioning generally reduces efficiency and maximum achievable
performance, but also it is very difficult to partition properly for parallel
tasks, while many emerging real-time workloads, such as deep neural network
(DNN)~\cite{Bojarski2016, nvidia2017bb8}, are highly parallel.

In this paper, we present RT-Gang: a novel real-time gang scheduling
framework that 
can \emph{efficiently} and \emph{predictably} utilize modern high-performance
multicore architectures for safety-critical real-time systems.  We focus on
supporting emerging \emph{parallel} real-time workloads, such as DNN-based
real-time sensing and control tasks~\cite{Bojarski2016,nvidia2017bb8}, while
also supporting legacy single-threaded real-time applications.
Our key observation is that, 
from the worst-case execution time (WCET) standpoint, \emph{scheduling one
parallel real-time task at a time} is better than co-scheduling multiple
parallel real-time tasks, because the latter case must assume highly
pessimistic WCETs on multicore architecture (see Section~\ref{sec:motivation}).

In RT-Gang, all threads of a parallel real-time task form a real-time
gang and the scheduler globally enforces a one-gang-at-a-time
scheduling policy to guarantee tight and accurate task WCET.
Assuming all real-time tasks are parallelized and each parallel task can fully
take advantage of all the available computing resources on the platform, this
one-gang-at-a-time approach essentially renders parallel real-time task
scheduling on multicore into the well-understood single-core real-time
scheduling problem~\cite{sha2004real}. The most significant benefit of this
transformation is that we no longer need to worry about shared resource
interference because all scheduled threads are part of a single real-time task
and that resource sharing is \emph{constructive} rather than destructive
in this setting. Assuming the WCET of a parallel real-time
task is estimated in isolation, RT-Gang guarantees that the WCET
will be respected regardless of other tasks on the system.
Furthermore, we can apply
well-understood single-core based real-time task scheduling policies and
analysis methodologies~\cite{lehoczky1989rate,sprunt1989aperiodic,Audsley93RTA}
without making strong assumptions on the underlying multicore
architectures.

Assuming all real-time tasks are perfectly parallelized is, however,
unrealistic. Many real-time tasks are difficult or impossible to 
parallelize. Also, parallelized code often does not scale well.
Therefore, our one-gang-at-a-time policy can significantly under-utilize
computing resources when imperfectly parallelized or single-threaded
real-time tasks are scheduled one-at-a-time.
We address the potential resource under-utilization problem in the
following two ways.

First, we introduce the notion of a \emph{virtual gang}. We define a virtual
gang task as a group of real-time tasks that are explicitly linked and scheduled
together as if they are threads of a single parallel real-time task under our
gang scheduler. Although real-time tasks in a virtual gang could 
interfere with each other, because the members of a gang are statically
determined at design time, analyzing their WCET impacts, while not
the focus of this paper, is easier under RT-Gang (e.g., via measurement).

Second, we allow co-scheduling of best-effort tasks on any of the available idle system
cores but with a condition that the cores are strictly regulated by a memory
bandwidth throttling mechanism~\cite{yun2017bwlock}. Each real-time gang defines
its tolerable maximum memory bandwidth, which is then strictly enforced by the
throttling mechanism to ensure bounded interference to the real-time gang task.

We implement RT-Gang in Linux kernel and evaluate it on two
representative embedded multicore platforms, NVIDIA Jetson TX-2 and
Raspberry Pi 3, using both synthetic and real-world workloads. The
results show that RT-Gang dramatically improves system
predictability while the measured overhead is negligible.

In summary, we make the following \textbf{contributions}:
\begin{itemize}
\item We propose RT-Gang, a novel gang scheduling framework that 
  enables analyzable parallel real-time scheduling on multicore
  platforms. 
\item We introduce the \emph{virtual gang} concept to minimize
  resource utilization by statically grouping multiple real-time
  tasks as a single schedulable real-time gang.
\item We further improve efficiency by integrating a state-of-the-art
  memory bandwidth throttling framework to enable safe best-effort
  task execution on any existing idle cores.
\item We implement RT-Gang in Linux kernel and present its evaluation
  results on two different embedded platforms using both synthetic and
  real-world workloads. We also provide RT-Gang as
  open-source~\cite{rt-gang}.
\end{itemize}

The rest of this paper is organized as follows: We present a motivating
case-study in Section~\ref{sec:motivation}.
In Section~\ref{sec:design}, we present the design of RT-Gang, and in
Section~\ref{sec:impl}, we describe its implementation details in Linux.
In Section~\ref{sec:eval}, we present our evaluation results.
After discussing potential use-cases and related work in
Section~\ref{sec:discussion} and Section~\ref{sec:related},
respectively, we conclude in Section~\ref{sec:conclusion}.
 \section{Motivation}\label{sec:motivation}

In this section, we provide evidence that shows why scheduling one real-time
gang at a time can be better from the perspective of task WCETs through a
case-study.

\begin{figure}[t]
  \centering
  \subfigure[Effect of DNN parallelization]{
    \includegraphics[width=\linewidth]{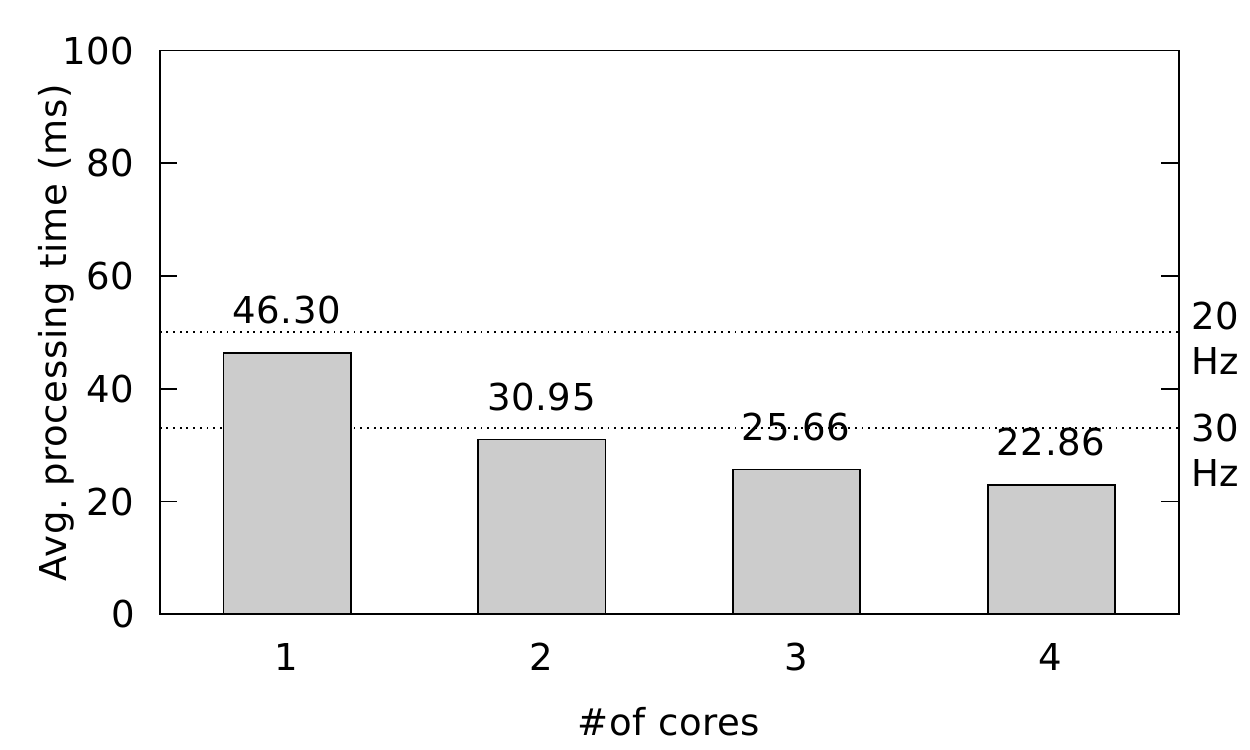}
    \label{fig:perf-vs-corecnt}
  }
  ~~~
  \subfigure[Effect of co-scheduling]{
    \includegraphics[width=\linewidth]{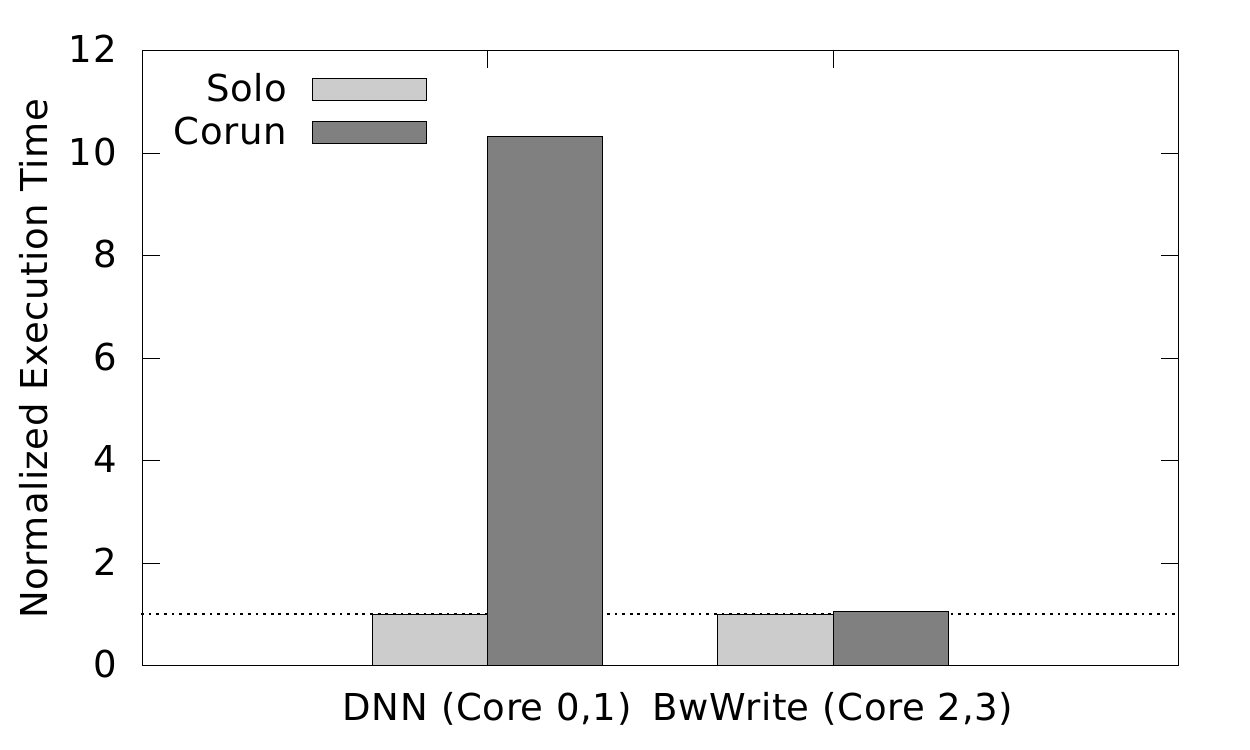}
    \label{fig:perf-cosched}
  }
  \caption{(a) Average control loop execution time vs. \#of CPU cores;
  (b) performance impact of co-scheduling (DNN on Core 0,1; BwWrite, a
    memory benchmark~\cite{valsan2016taming}, on Core 2,3)}
  \label{fig:perf-dnn}
\end{figure}

In this case-study, we use a deep neural network (DNN) based real-time control
task of DeepPicar~\cite{michael2018deep} as our workload. The control
loop uses a single deep neural network (DNN) to produce the car's 
steering angle control output from raw images of the car's front-facing camera
in real-time. Importantly, its DNN architecture is the same as the one used in
NVIDIA's real self-driving car called DAVE-2~\cite{Bojarski2016}.

Note that DNN processing is highly compute and data intensive, which is thus
often \emph{parallelized} to improve performance.
Figure~\ref{fig:perf-vs-corecnt} shows the average execution times of the
control loop while varying the number of CPU cores used
on a quad-core embedded platform (Raspberry Pi 3). It can be seen that as we
assign more cores for DNN processing, the performance improves---from 46.30 ms
on a single core to 22.86 ms on four cores. If the real-time requirement is
30Hz, one might want to parallelize the DNN using two cores, while
\emph{co-scheduling} other tasks on the other two remaining cores.

\begin{figure*}[t]
  \centering
  \includegraphics[width=.85\textwidth, height=4.5cm]{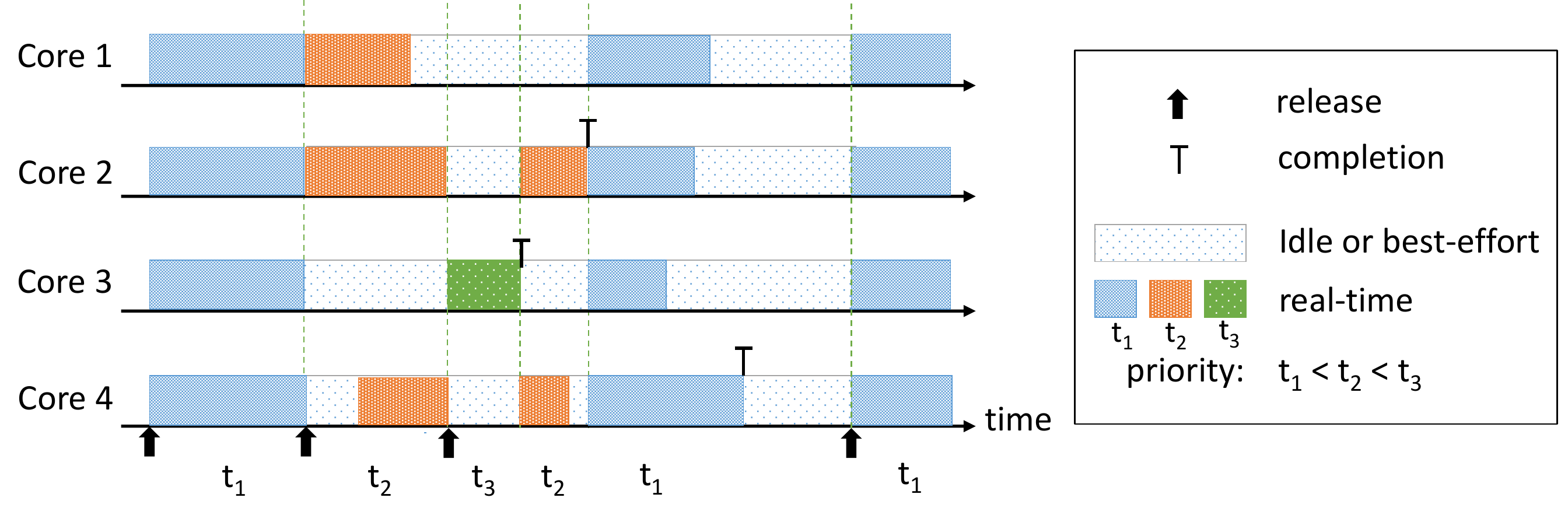}
  \caption{Proposed real-time gang scheduling approach}
  \label{fig:vscrm}
\end{figure*}

Figure~\ref{fig:perf-cosched} shows the timing impact of such co-scheduling,
where the DNN control task and a memory intensive task are scheduled first in
isolation (\emph{Solo}) and then together (\emph{Corun}).

The interesting, and troubling, observation is that the two tasks experience
dramatically different timing impact: the DNN control task suffers 10.33X
slowdown, while the memory benchmark suffers only 1.05X slowdown.

For safety-critical real-time systems, this means that extremely pessimistic
task WCETs must be assumed to be safe. The potential timing impact of
interference highly depends on task's memory access patterns and the underlying
hardware. For example, we observed more than 100X slowdown (two orders of
magnitudes) using a linked-list iteration task on the same computing platform
used above, \emph{even after} we partition core as well as the shared cache
among the tasks. Similar degrees of timing impacts of interference have been
reported in recent empirical studies on contemporary embedded multicore
platforms~\cite{valsan2017addressing,valsan2016taming,yun2015ospert,michael2018deep,michael2019dos}.

When a task's WCET has to be assumed 10X or 100X its solo execution time, we
can see why in aviation industry, it makes perfect sense to disable all but one
core~\cite{kim2017attacking} and why the certification authorities in US and
Europe recommend it for
certification~\cite{faa2016certification,faa2014certification}. However,
disabling cores obviously defeats the purpose of using multicore platforms in
the first place---the need of more performance.

In our DNN control-task case-study above, a better approach is to use all four
cores just for the parallelized DNN processing task---without
co-scheduling---which would result in quicker completion of the control task.
More importantly, because there would be no other competing co-scheduled
tasks, there's no need to pessimistically estimate the task's WCET.
This, in turn, will achieve better overall schedulability.

In this sense, from
the WCET standpoint, scheduling fully parallelized tasks one-at-a-time might be
better than co-scheduling multiple of them at the same time. Generally
applying this approach, however, has two major issues. First, not all
real-time tasks can be easily parallelized. Second, parallelization often does
not scale due to synchronization and other overheads. Therefore, the danger is
that some cores may be under-utilized under the one-at-a-time
scheduling policy.

In summary, we have made a case why, from the WCET standpoint, scheduling one
parallel real-time task at a time
can be better than co-scheduling multiple
parallel tasks simultaneously, although possible under-utilization of the
computing resources is a concern.
 \begin{figure*}[t]
  \centering
  \includegraphics[width=.8\textwidth, height=6.2cm]{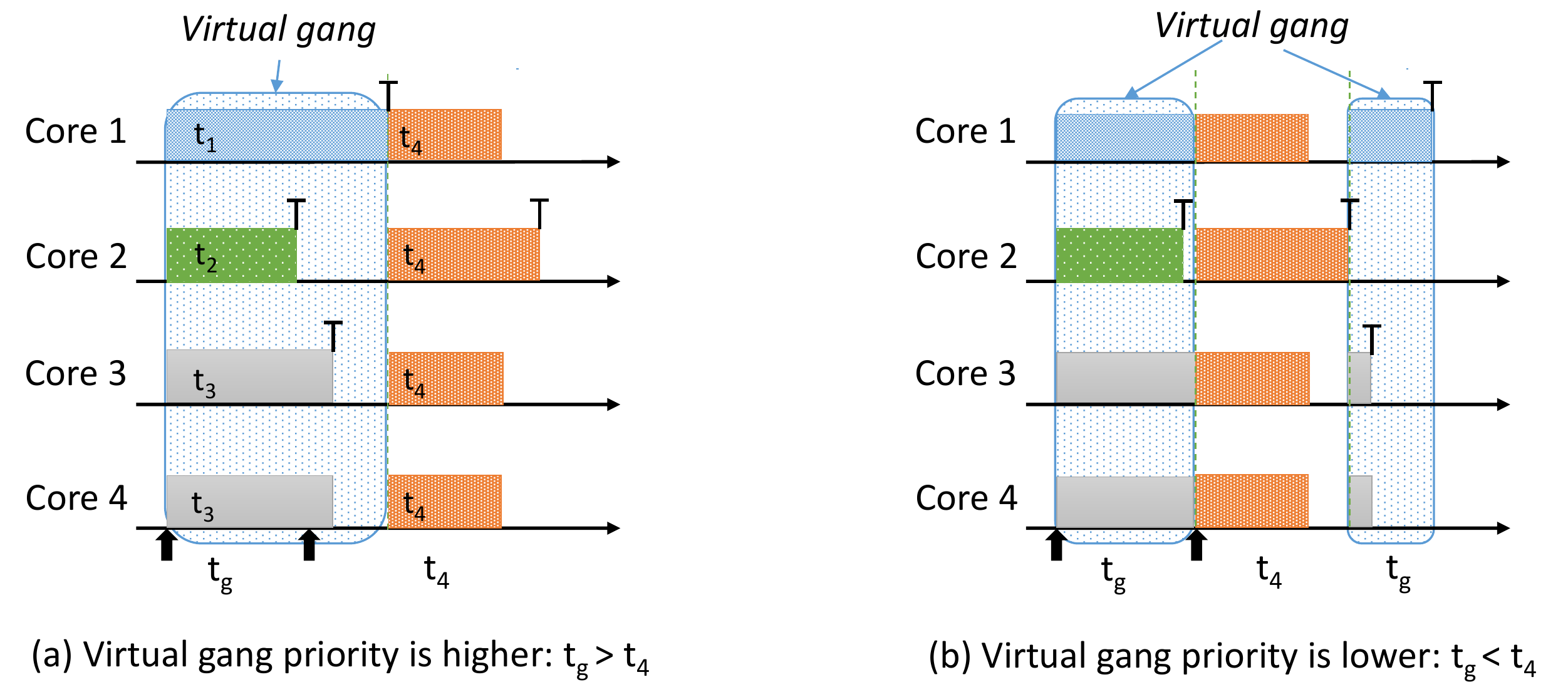}
  \caption{Virtual gang scheduling example}
  \label{fig:vgang}
\end{figure*}

\section{RT-Gang}\label{sec:design}
In this section, we describe the design of RT-Gang.

\subsection{Assumptions and Objectives}
We consider a shared memory based multicore processor. We assume a
hierarchical OS scheduling framework in which real-time tasks are strictly
prioritized over best-effort tasks (e.g., Linux).
We assume that each task is composed of
one or more threads. We assume that the thread to core assignment of
each parallel real-time task, but not best-effort ones, is given and
fixed (i.e., no migration).
We assume that the WCET of each real-time task 
is either experimentally measured or analytically computed
(e.g.,~\cite{skalistis2018safe}) in isolation.

Our objective is to eliminate the need to analyze
interference \emph{between} the real-time tasks on a multicore platform by turning
multicore parallel real-time scheduling into an equivalent of
single-core real-time task scheduling, while minimizing potential
utilization loss.

\subsection{Design Overview}
RT-Gang is based on a simple idea: \emph{schedule one real-time task---parallel
  or not---at a time.} When a real-time task is released, all of its
threads, called a \emph{gang}, shall be scheduled all at once---if the task
is the highest priority one---or 
not at all---if the task's priority is lower than the currently scheduled
real-time task---even if there exist some idle cores. In other words, we
implement a version of gang scheduler, but unlike prior gang scheduler
designs~\cite{goossens2010rtns_gangftp,kato2009rtss_gangedf,dong2017rtss_analysis},
we do not allow co-scheduling of other real-time tasks even when there
are available idle cores.
We do allow, however, co-scheduling best-effort tasks with strictly imposed
limits on their allowed memory access rates by integrating an
existing memory bandwidth throttling
mechanism~\cite{yun2017bwlock,ali2018protecting}. 

In our approach, each real-time task declares its maximum allowed interfering
memory traffic from the best-effort tasks on different cores, which is then
enforced by the OS at runtime for the cores that schedule best-effort tasks, if
such cores exist. In this way, the parallel real-time task, a
real-time gang, is guaranteed to be able to use all available
computing resources, and the maximum interference is strictly 
limited to a certain threshold value, determined by the task itself in
advance. If the real-time gang needs maximum isolation, it can set its
interference threshold value to be zero, preventing any co-scheduling
of best-effort tasks.

Figure~\ref{fig:vscrm} shows an example schedule under RT-Gang
framework. In this example, three real-time tasks $t_1$, $t_2$, and $t_3$ (in
increasing priority) are scheduled. The task $t_1$ has four threads, while
$t_2$ and $t_3$ have three threads and one thread, respectively.

At first, $t_1$ is scheduled. When $t_2$ is released, it preempts $t_1$ because
it has higher priority. Note that even if Core 3 and 4 are idle at the time,
$t_1$ cannot use the cores. When $t_3$, a single-threaded task, becomes ready,
all cores except Core 3 become idle to guarantee that $t_3$ is the only
real-time task in the entire system. In this way, our real-time gang scheduler
strictly enforces the one real-time gang at a time policy.

Note that the dot-filled rectangles are \emph{slack-durations} during which
best-effort tasks can be scheduled, using a general purpose scheduler, such as
Linux's Completely Fair Scheduler (CFS)~\cite{CFS}, but they will be throttled
based on each real-time task's declared tolerable threshold value.

RT-Gang's design approach offers several major benefits. First, by
scheduling one real-time gang 
at a time, we no longer need to worry about interference from other
real-time tasks. Possible interference from best-effort tasks is strictly
regulated by the threshold value determined by the task itself. Thus, we can
obtain accurate WCET of a real-time task (e.g., measure the timing
while injecting the threshold amount of memory traffic). Also, as shown
in~\cite{pellizzoni2016memory}, we can obtain better analytic memory
interference bounds when we control the amount of competing memory traffic.
In other words, we no longer need to deal with highly pessimistic 10X or 100X
WCETs. An equally desirable aspect of this approach is that it renders the
problem of scheduling parallel real-time tasks on multicore as the simple,
well-understood classical real-time scheduling problem on single-core
systems~\cite{lehoczky1989rate,sprunt1989aperiodic}. Thus, we can directly
apply classical single-core analysis methods~\cite{Audsley93RTA}.

\subsection{Virtual Gang}
Our basic real-time gang scheduling approach described above would work well
when all real-time tasks are perfectly parallelized, but it may under-utilize
the system if real-time tasks are single-threaded and best-effort tasks are not
available. Furthermore, some of them may miss the deadlines, while they might
be schedulable if all cores are used.

We propose the \emph{virtual gang} concept  to mitigate the resource
under-utilization problem of imperfectly parallelized real-time tasks under our
gang scheduling policy. We define a virtual gang as a group of real-time
tasks that are explicitly linked and scheduled together as if they are threads
of a single real-time task under our gang scheduler. A virtual gang task has a
fixed priority value that applies to all of its members. From the gang
scheduler's point of view, a virtual gang is just like a normal real-time task,
except the scheduled threads are not from the same task but from multiple
different tasks. The composition of a virtual
gang must be explicitly determined by the system designer at \emph{design
time}. Once a real-time task becomes a member of a virtual gang, all
of its threads are scheduled simultaneously with the threads of
the other member real-time tasks of the virtual gang.

Figure~\ref{fig:vgang} shows two example schedules involving a virtual gang
$t_g$, which consists of three separate real-time tasks: $t_1, t_2$ and $t_3$.
Note that $t_1$ and $t_2$ are single-threaded while $t_3$ is multi-threaded but
only two threads are used. All real-time tasks in the virtual gang $t_g$ share
the same priority. Therefore, from the scheduler's point-of-view, all threads
of the three tasks are treated as a single (virtual) real-time task $t_g$, the
virtual gang task. For instance, in inset (a) of Figure~\ref{fig:vgang}, a
newly arrived task $t_4$ cannot be scheduled until $t_1$'s last
thread is completed because $t_g$'s priority is higher than $t_4$. In inset
(b), on the other hand, $t_4$ immediately preempts all active threads of the
virtual gang $t_g$, because $t_g$'s priority is lower than $t_4$'s.

Although the real-time tasks in a virtual gang can destructively interfere with
each other, their membership is explicitly determined at the design time.
Therefore, the effect of shared resource interference among the tasks in a
virtual gang can be carefully analyzed, either empirically or analytically, and
taken into account for system-level schedulability analysis.

\subsection{Safe Best-Effort Task Co-Scheduling}
Because our real-time gang scheduling approach strictly disallows
concurrent real-time tasks, which are not part of the currently
scheduled real-time gang, it is possible that some cores may be idle.
As we discussed earlier, we allow scheduling of best-effort tasks on
those idle cores with a condition that their interference is strictly
bounded by integrating a memory bandwidth throttling mechanism as
found in~\cite{yun2017bwlock}.

The throttling mechanism in ~\cite{yun2017bwlock} uses per-core hardware
performance counters to bound the maximum number of memory transactions within
a given time interval (e.g., \texttt{1-msec} period) to a certain programmable
threshold (budget). When the core reaches the programmed threshold, the
hardware counter generates an overflow interrupt, which is then handled by the
OS to stop the core until the next time period begins.

Assuming that the currently scheduled real-time gang is actively using $k$
cores out of $m$ cores ($k \leq m$), there are $m - k$ idle cores on which we
can schedule best-effort tasks---i.e., those that do not have hard real-time
requirements. The best-effort tasks scheduled on the idle cores are given
strict limits in terms of their memory bandwidth usage so that their impact to
the real-time gang is bounded. The bandwidth limit of the best-effort tasks is
determined by the currently scheduled real-time gang in the system. When the
real-time gang is being scheduled on $k$ cores, all the $m-k$ cores are
throttled according to the bandwidth threshold of the gang.

\subsection{Illustrative Example}\label{sec:example}
In this subsection, we provide an illustrative example to compare
scheduling under RT-Gang with a traditional co-scheduling approach.

\begin{figure}[t]
\centering
\subfigure[Example schedule of a co-scheduling scheme (w/o interference)]{
	\includegraphics[width=\linewidth]{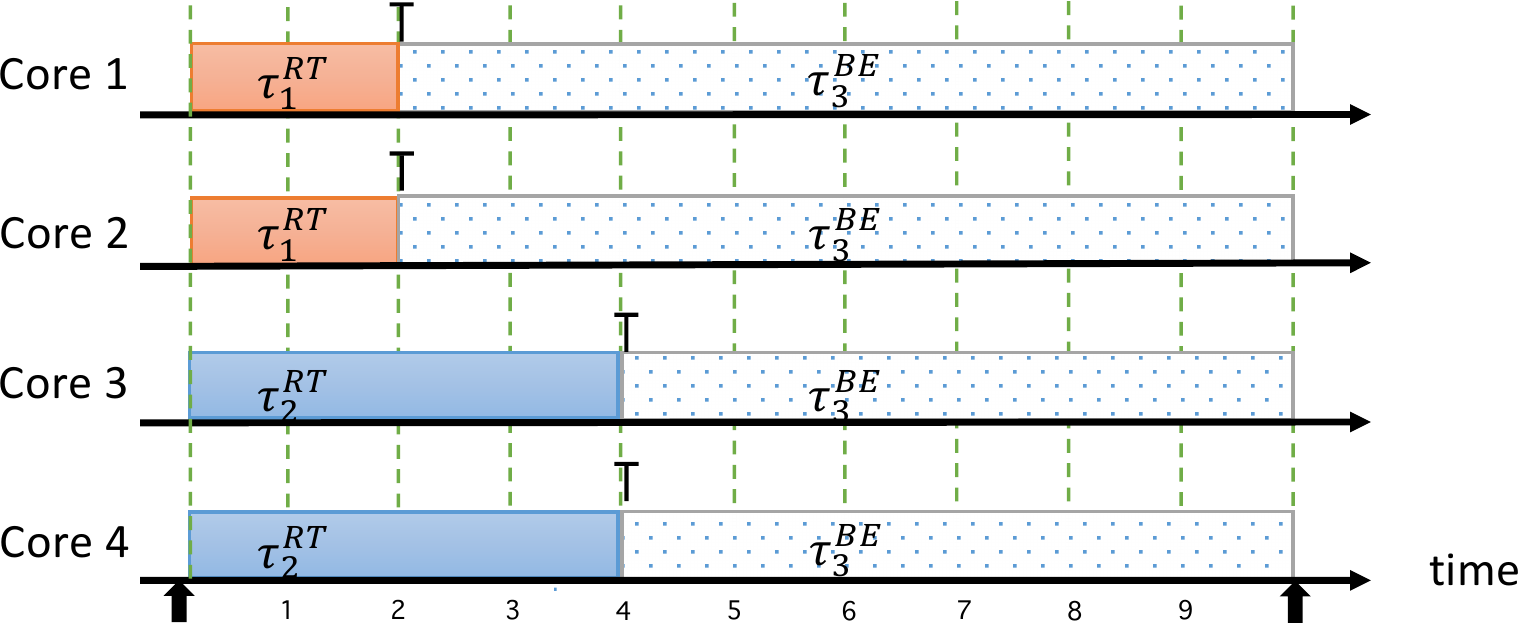}
	\label{fig:ex-cosched}
}
~~
\subfigure[Example schedule of RT-Gang ]{
	\includegraphics[width=\linewidth]{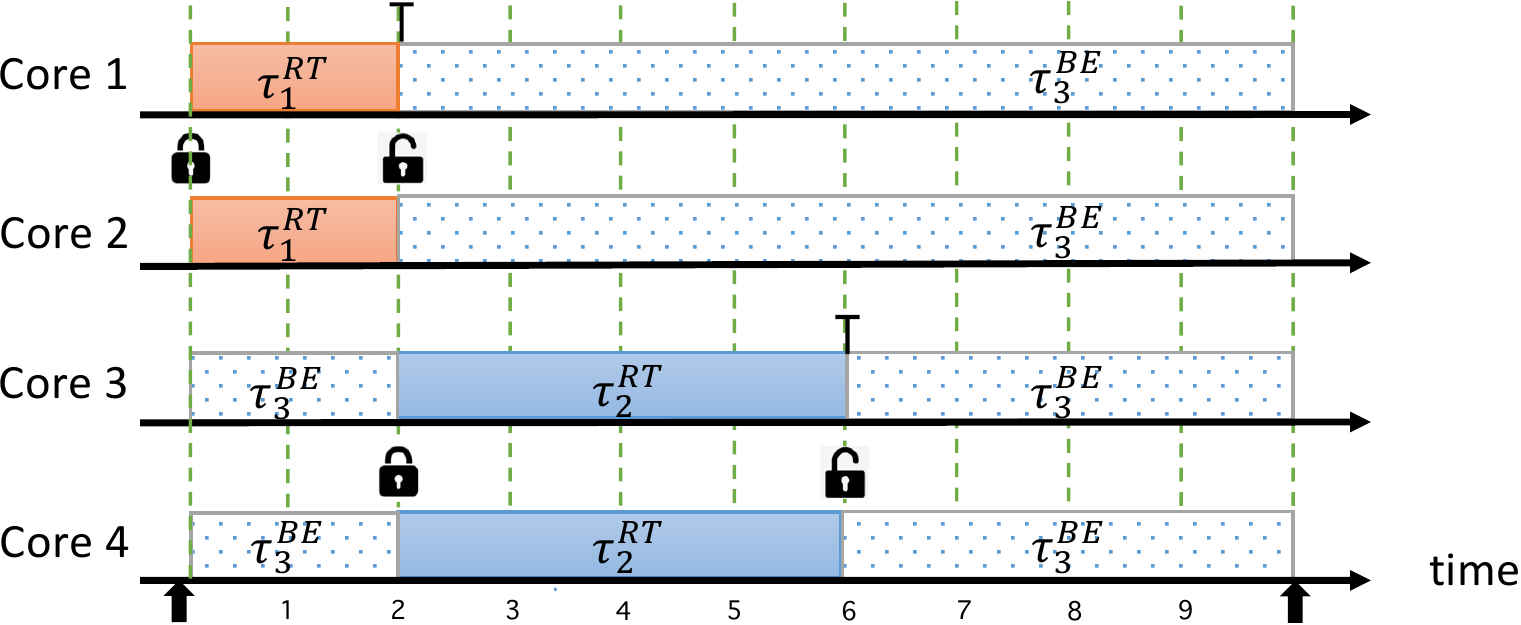}
	\label{fig:ex-gsched}
}
~~
\subfigure[Example schedule of a co-scheduling scheme (with interference)]{
	\includegraphics[width=\linewidth]{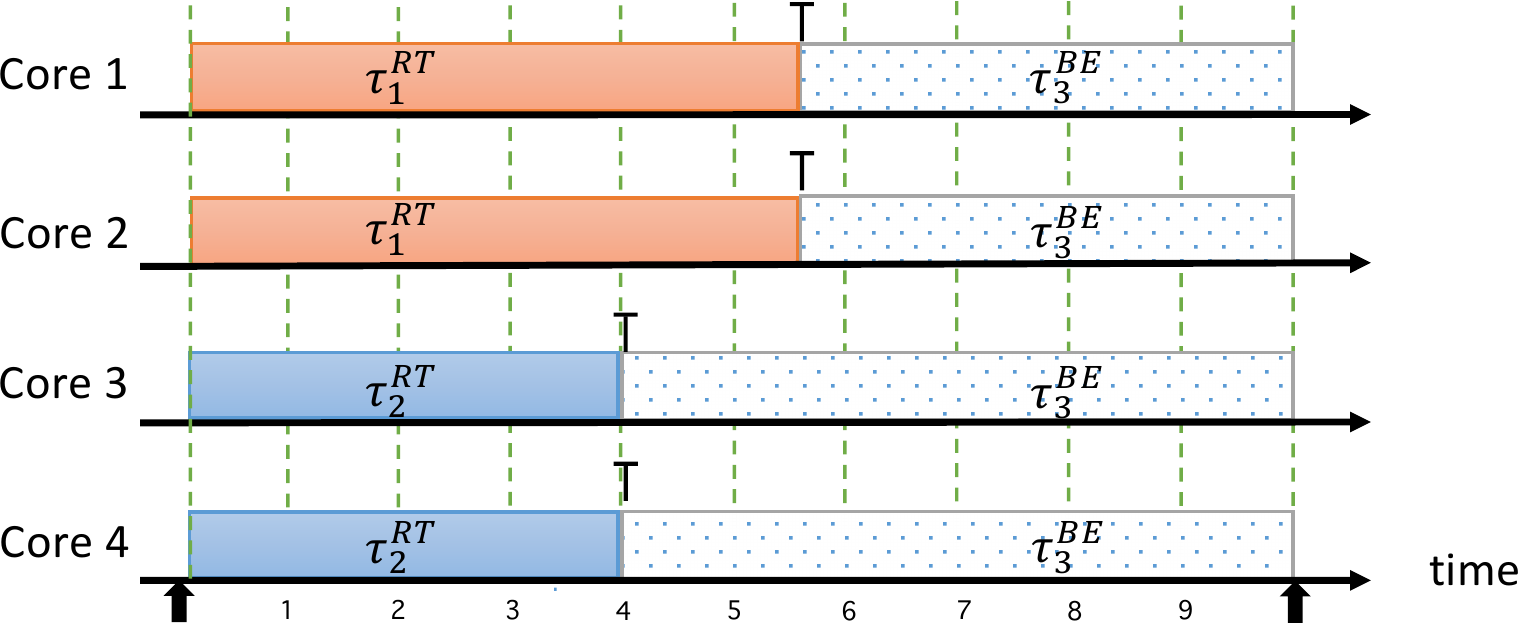}
	\label{fig:ex-coschedInter}
}
\caption{Example schedules under different scheduling schemes}
\label{fig:example}
\end{figure}

{
\renewcommand{\arraystretch}{1.5}
\begin{table}[t]
	\centering
	\begin{tabularx}{0.7\linewidth}{l|l|l|l}
		\toprule
		Task 			& WCET ($C$)	& Period ($P$)         & \# Thread\\
		\midrule
		$\tau_{1}^{RT}$		& 2 			& 10		& 2\\
		$\tau_{2}^{RT}$		& 4 			& 10		& 2\\
		$\tau_{3}^{BE}$		& $\infty$		& N/A		& 4\\
		\bottomrule
	\end{tabularx}
	\caption{Taskset parameters of the illustrative example}
	\label{tbl:example}
\end{table}
}

Let us consider a multicore system with four homogeneous CPU cores. We want to
schedule a parallel taskset, shown in Table~\ref{tbl:example}, in the system.
$\tau_{1}^{RT}$ and $\tau_{2}^{RT}$ are real-time tasks. $\tau_{3}^{BE}$ is a
best-effort task, which is scheduled under the default CFS scheduler. For the
sake of simplicity, we assume that all threads of a task have the same
compute time and are released simultaneously at the start of the period.
We assume that all threads are statically pinned to specific CPU cores and they
do not migrate during their execution. The OS scheduler tick interval is
assumed to be \texttt{1-msec}.

Figure~\ref{fig:ex-cosched} shows the scheduling timeline under a traditional
co-scheduling approach. For this figure, we assume that tasks on
different cores do not suffer interference due to contention in shared
hardware resources. Under co-scheduling, $\tau_{1}^{RT}$
completes its execution at \texttt{2-msec}. This leaves \texttt{8-msec}
\emph{slack duration} on the two cores on which this task was executing.
Similarly, $\tau_{2}^{RT}$ leaves a slack duration of \texttt{6-msec} on its
cores. Considering the system as a whole, the total slack duration left in this
schedule is \texttt{28-msec}. The slack duration can be used to
schedule the best-effort task $\tau_{3}^{BE}$.

Figure~\ref{fig:ex-gsched} shows the scheduling timeline under
RT-Gang. Under this schedule, 
$\tau_{1}^{RT}$ gets to run first. While $\tau_{1}^{RT}$ is executing,
$\tau_{2}^{RT}$ is blocked thanks to our one-gang-at-a-time policy.
Once $\tau_{1}^{RT}$ finishes
its execution at \texttt{2-msec} mark, $\tau_{2}^{RT}$ starts executing and
completes at \texttt{6-msec} mark. Under this scheme, the total slack duration
left in the system is again \texttt{28-msec}, assuming that $\tau_3^{BE}$ is
not throttled (i.e., its memory bandwidth usage is less than the
allowed threshold).

Now we consider the case where the real-time tasks can destructively interfere
with each other due to resource contention. Borrowing
from the DNN case-study introduced in Section~\ref{sec:motivation}, we assume
that the execution time of $\tau_{1}^{RT}$ increases 10X, when it is
co-scheduled with $\tau_{2}^{RT}$. $\tau_{2}^{RT}$, on the other hand, stays
unaffected under co-scheduling. Figure~\ref{fig:ex-coschedInter} shows the
scheduling timeline for this case under co-scheduling scheme. As can be seen
from the figure, while $\tau_{2}^{RT}$ is executing, the progress of
$\tau_{1}^{RT}$ would be slowed due to interference. 
At \texttt{4-msec} mark when
$\tau_{2}^{RT}$ finishes its execution, $\tau_{1}^{RT}$ has only made
\texttt{20\%} progress and it still has \texttt{1.6-msec} of its original
compute time left. For this reason, $\tau_{1}^{RT}$ completes its execution at
\texttt{5.6-msec}. In this case, the total slack duration for
best-effort tasks is \texttt{20.8-msec}.

Under RT-Gang, the scheduling timeline of the real-time tasks remains
{\bf the same} as the one shown in Figure~\ref{fig:ex-gsched} because
$\tau_{1}^{RT}$ and $\tau_{2}^{RT}$ can never run at the same time. In
other words, regardless of task and hardware characteristics,
real-time tasks' execution times would remain the same. 
The slack duration remains unchanged as well, which can be utilized
for scheduling best-effort tasks although they are strictly regulated
with a memory bandwidth throttling mechanism, shown as the ``locked''
duration in Figure~\ref{fig:ex-gsched}.

In summary, the major benefits of RT-Gang are: (1)  Parallel real-time tasks enjoy highly \emph{predictable} task WCETs
  regardless of the characteristics of scheduled tasks and the
  underlying hardware because only one real-time task is scheduled at a time;
(2)  Best-effort tasks can be \emph{safely} scheduled on any idle
  cores because the integrated throttling mechanism guarantees bounded
  interference.

 \section{Implementation}\label{sec:impl}
In this section, we describe the implementation details of RT-Gang in
Linux kernel.

\SetKwProg{funct}{function}{}{}
\SetKwFor{ForLCore}{for\_each\_locked\_core}{}{}
\SetKwFor{For}{for\_each\_sched\_class}{}{}
\DontPrintSemicolon

\begin{algorithm}[t]
\SetKwFunction{name}{glock}
\SetKwProg{type}{struct}{}{}
\SetKwFunction{func}{\_\_schedule}
\SetKwProg{algo}{void}{}{}
\SetKwFunction{funcA}{$^\ast$pick\_next\_task\_rt}
\SetKwProg{algoA}{task\_struct\_t}{}{}

\type{\name}{
\begin{tabularx}{\linewidth}{l>{$}c<{$}X}
 spinlock\_t			&&lock;\\
 bool				&&held\_flag;\\
 bitmask			&&locked\_cores;\\
 bitmask			&&blocked\_cores;\\
 task\_struct\_t$^\ast$		&&leader;\\
 task\_struct\_t$^\ast$		&&gthreads [NR\_CPUS];
\end{tabularx}
}

\nonl\;
\funct{\func{task\_struct\_t $^\ast$prev}}{
	\For{(class)}{
 		next = class$\rightarrow$pick\_next\_task (prev);

	\If{(next)}{
		context\_switch (prev, next);
	}
}
}
\KwRet

\nonl\;
\funct{\funcA{task\_struct\_t $^\ast$prev}}{
	spin\_lock (glock$\rightarrow$lock);\nonl\;
	\If {(glock$\rightarrow$held\_flag)}{
		try\_glock\_release (prev);
	}
	next =  rt\_queue [this\_cpu]$\rightarrow$next\_task;\nonl\;
	\uIf {(glock$\rightarrow$held\_flag == false)}{
		acquire\_gang\_lock (next);
	}
	\uElseIf {(next$\rightarrow$prio == glock$\rightarrow$leader$\rightarrow$prio)}{
		set\_bit (this\_cpu, glock$\rightarrow$locked\_cores);\nonl\;
		glock$\rightarrow$gthreads [this\_cpu] = next;
	}
	\uElseIf {(next$\rightarrow$prio $>$ glock$\rightarrow$leader$\rightarrow$prio)}{
		do\_gang\_preemption ();\nonl\;
		acquire\_gang\_lock (next);
	}
	\Else {
		set\_bit (this\_cpu, glock$\rightarrow$blocked\_cores);\nonl\;
		next = null;
	}
	spin\_unlock (glock$\rightarrow$lock);\nonl\;
}
\KwRet{next}

\caption{RT-Gang in Linux}
\label{alg:gsched}
\end{algorithm}

Algorithm~\ref{alg:gsched} shows the high-level pseudo-code of RT-Gang. The
implementation revolves around a data-structure \texttt{struct glock}
declared in \emph{Line-2} of the algorithm. This data-structure is
used for the following main purposes:
\begin{itemize}
\item Quickly check whether the gang scheduling lock is currently
  being held (\emph{held\_flag})
\item Track the cores, which are currently running real-time threads
  using a bitmask (\emph{locked\_cores})
\item Track the blocked cores, which have real-time tasks in
  their queues but cannot execute them due to the gang scheduling
  policy (\emph{blocked\_cores})
\end{itemize}

To simplify the implementation complexity, we assume that each
real-time gang in our system has a distinct real-time priority
value. In the following sections, we explain the main parts of the
algorithm in detail. 

\subsection{Gang Lock Acquisition}
Before a real-time task can get scheduled, it needs to acquire the gang
scheduling lock. Algorithm~\ref{alg:gschedAcq} shows the pseudo-code
of the lock acquisition function.
In this function, the gang-scheduling lock is marked as held by
setting the flag in the global \texttt{glock} data-structure.
The CPU core on which this function is invoked, is marked ``locked''
by setting its bit inside the \texttt{locked\_cores} bitmask.
The task that acquires the gang lock is marked as the gang-leader and
its task pointer is tracked inside an array, which is later used at
the time of lock release.

\begin{algorithm}[b]
\SetKwFunction{aglck}{acquire\_gang\_lock}

\funct{\aglck{task\_struct\_t $^\ast$next}}{
	glock$\rightarrow$held\_flag = true;\nonl\;
	set\_bit (this\_cpu, glock$\rightarrow$locked\_cores);\nonl\;
	glock$\rightarrow$gthreads [this\_cpu] = next;\nonl\;
	glock$\rightarrow$leader = next;\nonl\;
}
\KwRet

\caption{Gang Scheduling: Lock Acquisition}
\label{alg:gschedAcq}
\end{algorithm}

\subsection{Gang Lock Release}
This function is called to release the gang-scheduling lock on behalf of a task
which is going out of schedule. The pseudo-code of this function is shown in
Algorithm~\ref{alg:gschedRel}. Upon entering this function, it is checked
whether the thread going out of execution is part of the currently executing
gang. If this condition is true, the current CPU core is marked as
unlocked, by clearing its bit from the \texttt{locked\_cores} bitmask.

The next condition checked in this function is to make sure if all of the
threads belonging to current gang have finished their execution, which would
imply that the gang-lock is now free. This is done by checking if the
\texttt{locked\_cores} bitmask is zero. If this is the case, the
gang-scheduling lock is marked free and rescheduling
inter-processor interrupt (IPI) is sent to the CPU cores, which have
blocked real-time tasks in their ready queues by using the
\texttt{blocked\_cores} bitmask.

\begin{algorithm}[h]
\SetKwFunction{tgrel}{try\_glock\_release}

\funct{\tgrel{task\_struct\_t $^\ast$prev}}{
	\ForLCore {(cpu, glock$\rightarrow$locked\_cores)}{
		\If {(gthreads [cpu] == prev)}{
			clear\_bit (cpu, glock$\rightarrow$locked\_cores);\nonl\;
			gthreads [cpu] = null;\nonl\;

			\If {(mask\_is\_zero (glock$\rightarrow$locked\_cores))}{
				glock$\rightarrow$held\_flag = false;\nonl\;
				reschedule\_cpus (glock$\rightarrow$blocked\_cores);\nonl\;
				clear\_mask (glock$\rightarrow$blocked\_cores);\nonl\;
			}
		}
	}
}
\KwRet

\caption{Gang Scheduling: Lock Release}
\label{alg:gschedRel}
\end{algorithm}

\subsection{Gang Preemption}
The purpose of this function is to preempt all threads, which are part of the
currently running gang, so that a new higher priority gang may start its
execution. The pseudo-code of this function is shown in
Algorithm~\ref{alg:gschedPre}. In this function, the \texttt{locked\_cores}
bitmask is traversed to send rescheduling IPIs to all the
cores, which are currently marked as locked. Once this is done, the
\texttt{locked\_cores} bitmask is cleared and the threads being
tracked in the \texttt{gthreads} array are removed.

\begin{algorithm}[h]
\SetKwFunction{dgpre}{do\_gang\_preemption}

\funct{\dgpre{}}{
	\ForLCore {(cpu, glock$\rightarrow$locked\_cores)}{
		gthreads [cpu] = null;\nonl\;
	}
	reschedule\_cpus (glock$\rightarrow$locked\_cores);\nonl\;
	clear\_mask (glock$\rightarrow$locked\_cores);\nonl\;
}
\KwRet

\caption{Gang Scheduling: Gang Preemption}
\label{alg:gschedPre}
\end{algorithm}

\subsection{Main Gang Scheduling Algorithm}
The gang-scheduling algorithm resides in the critical path of the main
scheduler entry point function (\texttt{\_\_schedule}) in Linux and it
works by modifying the task selection heuristics of the real-time
scheduling class.

Algorithm~\ref{alg:gsched} shows the main scheduling function of
RT-Gang. The algorithm starts by checking whether gang-scheduling lock
is currently being held by any real-time task. If that is the case, the 
algorithm tries to release the gang-scheduling lock on behalf of the
\texttt{prev} task which is going out of schedule (\emph{Line-11}).

If the gang-scheduling lock is currently free, the algorithm acquires
the lock on the current core on behalf of the \texttt{next} real-time
task (\emph{Line-13}).

If the lock is currently not free, the algorithm checks whether the
\texttt{next} real-time task on this CPU core belongs to the same gang that is
holding the gang-scheduling lock (\emph{Line-14}). Here, we exploit the fact
that each real-time gang has a distinct priority as mentioned
earlier this section. Hence, if the \texttt{next} 
task and the gang leader task have the same real-time priority value, they are
considered a part of the same gang. In this case, the current core is marked
``locked'' by setting its bit in the \texttt{locked\_cores} bitmask
(\emph{Line-15}).

If the lock is not free and the \texttt{next} real-time task does not belong to
the currently executing gang, it is checked whether this task has a higher
priority than the gang in execution (\emph{Line-16}). If that is the case, the
currently executing gang is preempted and the gang-scheduling lock is acquired
on behalf of the \texttt{next} task (\emph{Line-17}).

If all of the above conditions fail---i.e., the gang-scheduling lock
is not free, the \texttt{next} real-time task does not belong to the
currently executing gang, and it is of lower priority, then the
\texttt{next} task is deemed ineligible for scheduling. In this case,
the current CPU core is marked as 
blocked by setting its bit in the \texttt{blocked\_cores} bitmask
(\emph{Line-19}) and the \texttt{next} task pointer is set to null so that no
real-time task is returned to the scheduler by the real-time
scheduling class.

Finally, the spinlock is released (\emph{Line-20}) and control is returned to
the scheduler (\emph{Line-21}); to either schedule the next real-time task (if
any) or go to the next scheduling class (CFS) to pick a best-effort task.

\subsection{Creation of Virtual Gangs}
In our implementation of RT-Gang in Linux, creation of \emph{virtual
gang} tasks is straight-forward. Because each real-time gang task has
a distinct real-time priority in our system, the only thing that a 
system programmer has to do to mark different real-time tasks as part of the
same virtual gang, is to assign them the same real-time priority value under
Linux. Once this is done, RT-Gang framework allows simultaneous execution of
such tasks just like real gangs.

\begin{figure*}[h]
\centering
\subfigure[without RT-Gang (Baseline Linux)]{
	\includegraphics[width=0.99\textwidth]{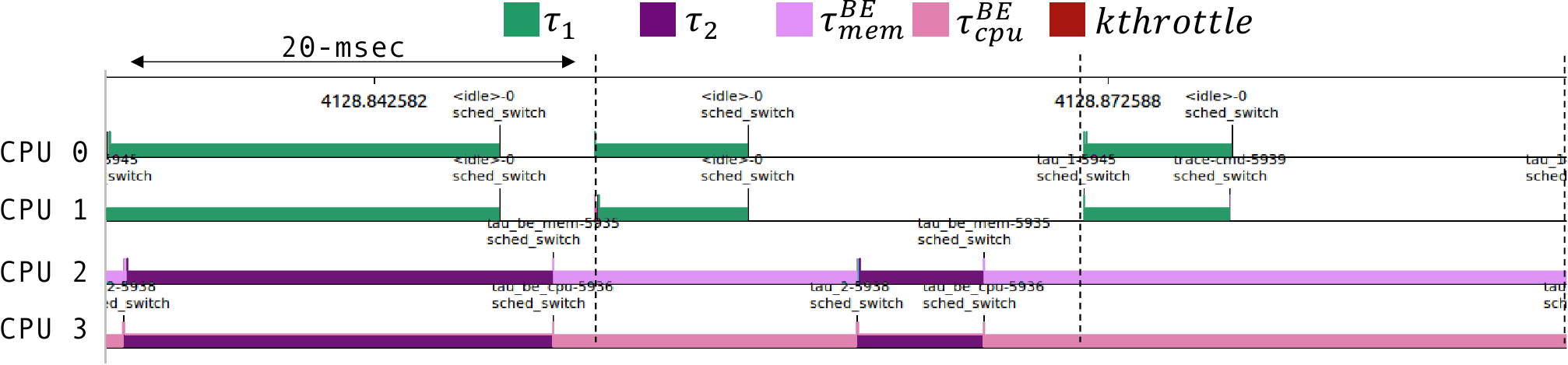}
	\label{fig:ev-ulock}
}
\subfigure[with RT-Gang]{
	\includegraphics[width=0.99\textwidth]{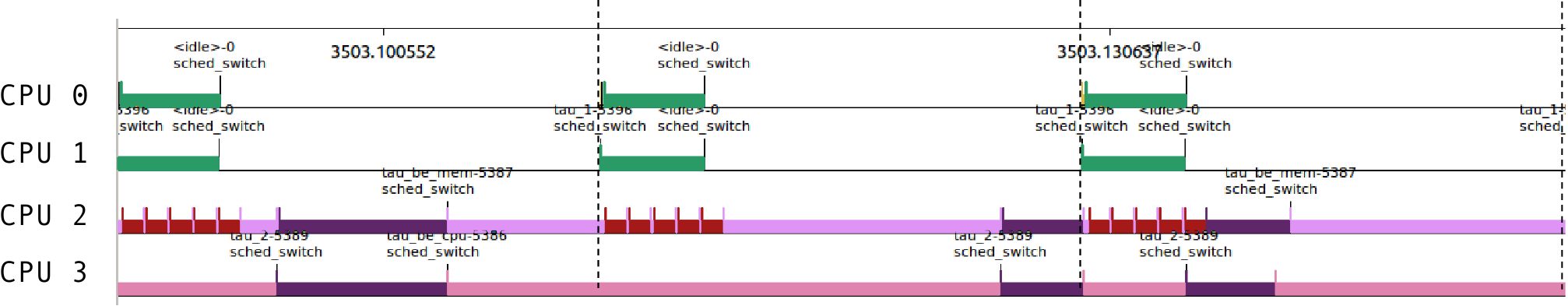}
	\label{fig:ev-lock}
}
\caption{Task execution traces. $\tau_1 (C_1=3.5, P_1=20)$,
  $\tau_2 (C_2=6.5, P_2=30)$: parallel RT tasks (2 threads / task);
  $\tau_{mem}^{BE}$, $\tau_{cpu}^{BE}$: memory and cpu intensive
  best-effort tasks respectively; 
  \emph{kthrottle}: injected kernel thread for throttling}
\label{fig:eval-gsched}
\end{figure*}

\subsection{Memory Bandwidth Throttling of Best-Effort Tasks}
We incorporated a memory bandwidth throttling framework based
on~\cite{yun2017bwlock} in RT-Gang. This framework provides system programmers
with a system-call; to mark the entire duration of a real-time application as
memory bandwidth sensitive (i.e., Coarse-Lock).  We update this system call
such that instead of providing a binary value to start/stop throttling, the
caller provides the acceptable memory threshold value for the real-time gang in
execution. This value is stored in the task structure of the RT-Gang leader as
an integer. We also modify the framework in~\cite{yun2017bwlock} such that in
every regulated interval, the memory bandwidth threshold value of the executing
gang is automatically enforced on all CPU cores executing best-effort tasks. In
this manner, we ensure that the real-time gang is protected from unbounded
interference from best-effort tasks.

\section{Evaluation}\label{sec:eval}
In this section, we present the evaluation results of RT-Gang.

\subsection{Setup}
We evaluate RT-Gang on two embedded platforms: Raspberry Pi3 and
NVIDIA Jetson TX2. Raspberry Pi3 is equipped with a Cortex-A53 based
quad-core CPU, which is representative of an energy 
efficient low-cost embedded multicore processor, while NVIDIA Jetson
TX2 is equipped with a six-core heterogeneous CPU (4X Cortex-A57 and
2X Denver~\footnote{We do not use the Denver complex in our
  experiments due to its lack of hardware counter support needed to
  implement throttling mechanism~\cite{ali2018protecting}}), which is representative of a high-end embedded processor.

On both platforms, we use Linux 4.4 kernel and implement RT-Gang by
modifying its real-time scheduler (\texttt{kernel/sched/rt.c}). Our
modification is about 500 lines of architecture neutral C code.

In all our experiments, we place the platform in the maximum performance mode
and disable the dynamic frequency scaling of CPU cores. We also shutdown the
graphical user interface and disable networking to minimize run to run variation
in the experiments.

\begin{figure*}[h]
\centering
\subfigure[TX2: DNN (2 Core) ]{
	\includegraphics[width=0.315\textwidth]{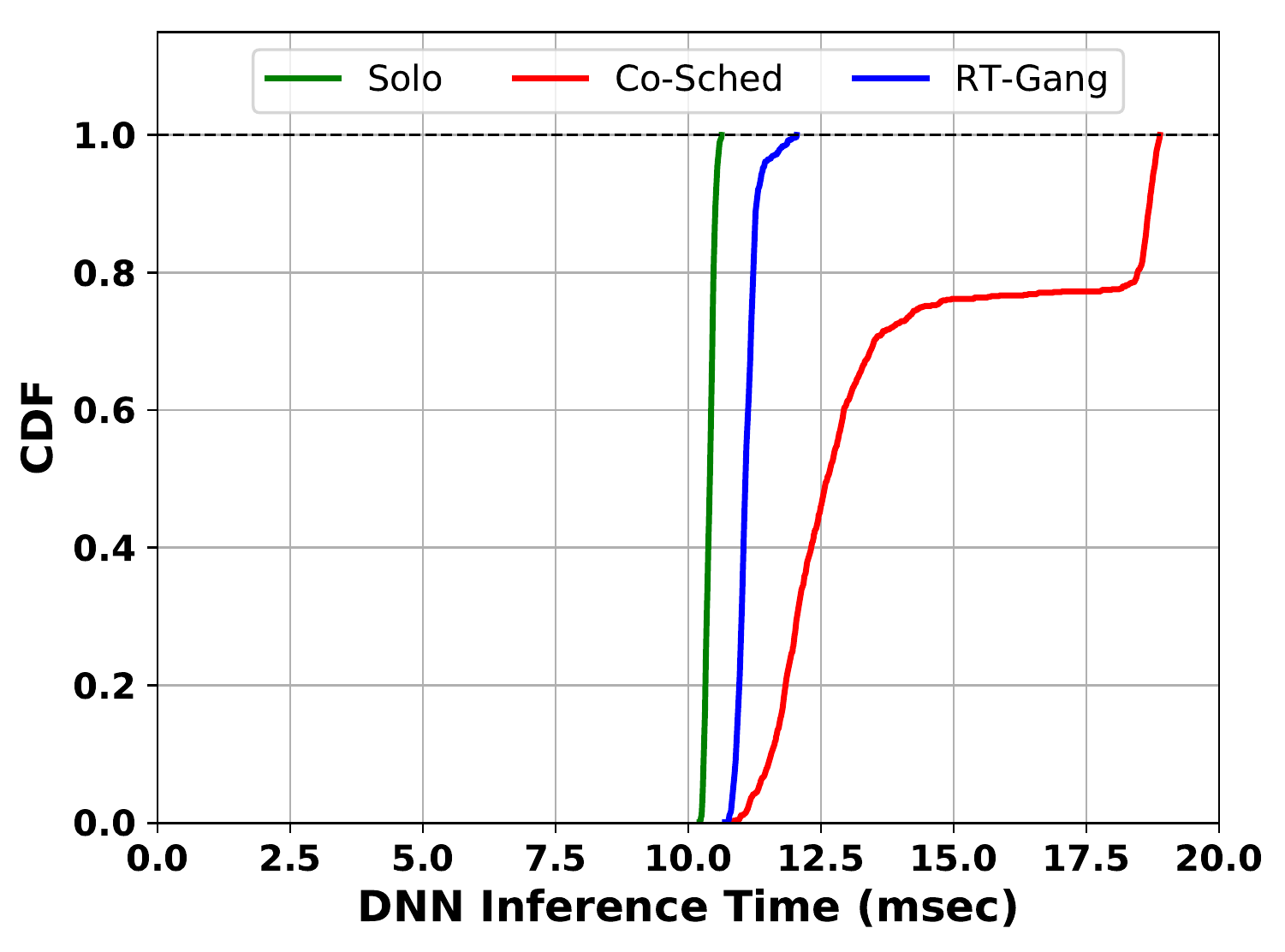}
	\label{fig:ev-tx_1n1c}
}
\subfigure[TX2: DNN (3 Cores)]{
	\includegraphics[width=0.315\textwidth]{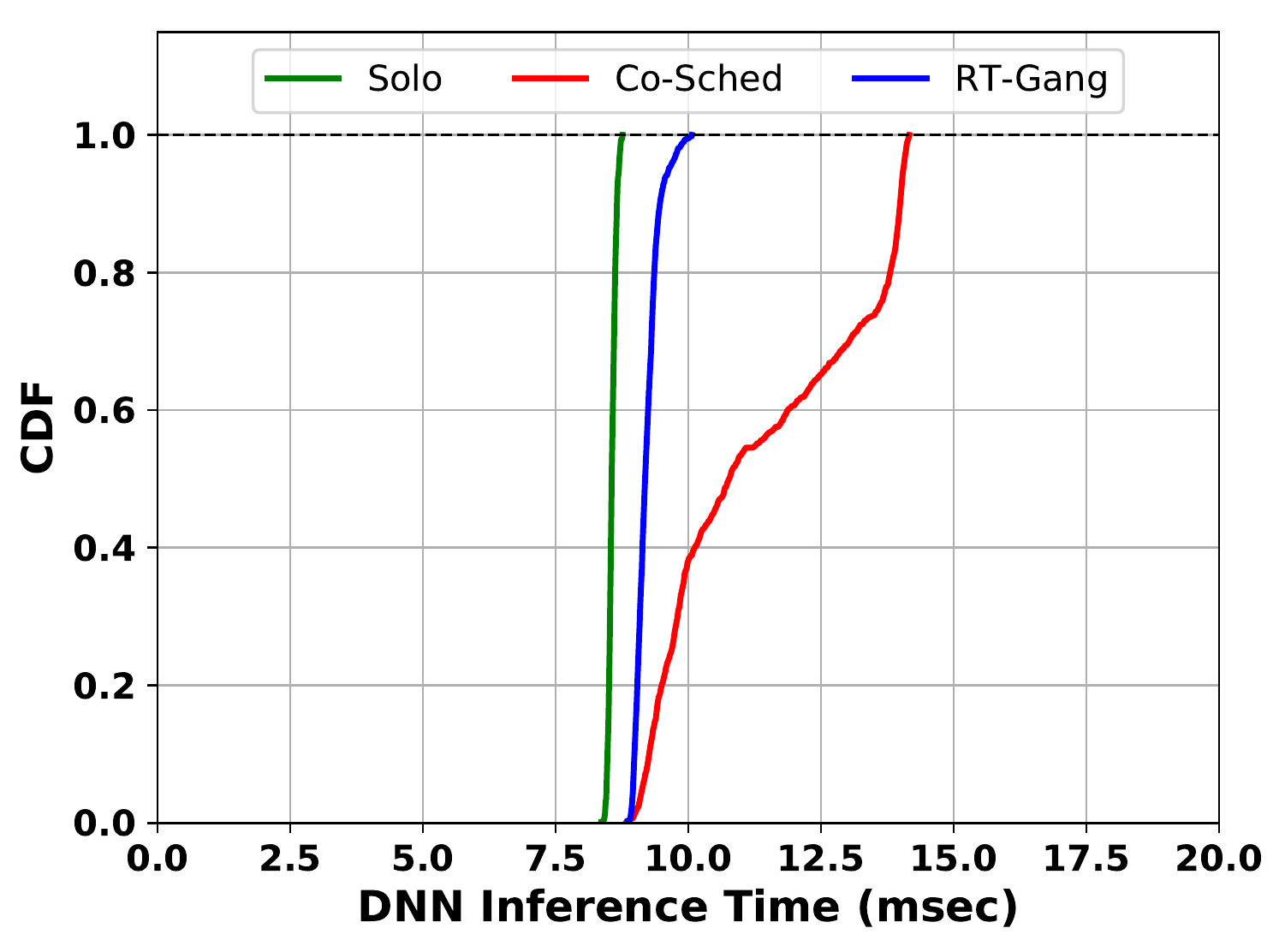}
	\label{fig:ev-tx_1n2c}
}
\subfigure[TX2: DNN (4 Cores)]{
	\includegraphics[width=0.315\textwidth]{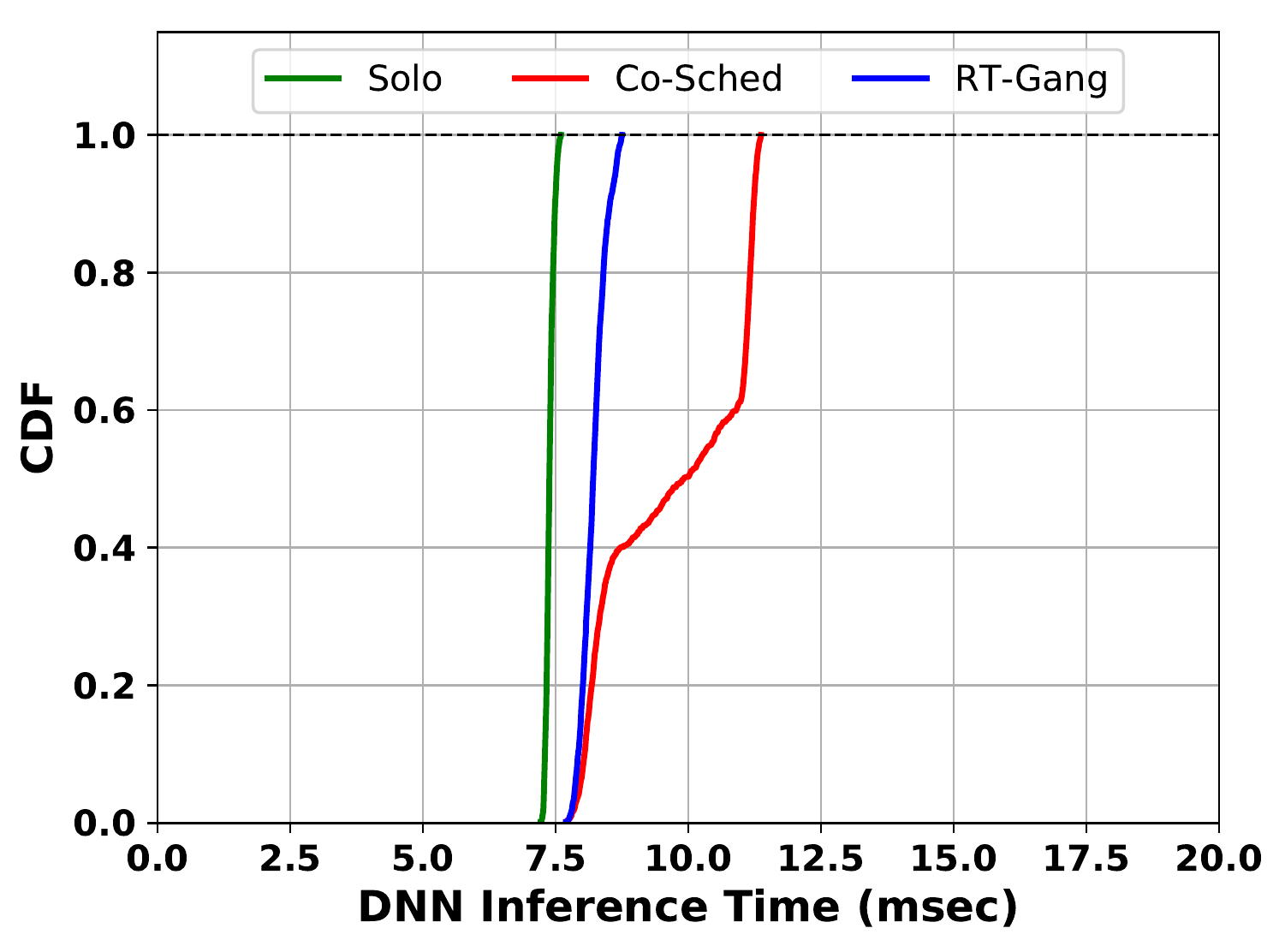}
	\label{fig:ev-tx_1n3c}
}
~~
\subfigure[Pi3: DNN (2 Core)]{
	\includegraphics[width=0.315\textwidth]{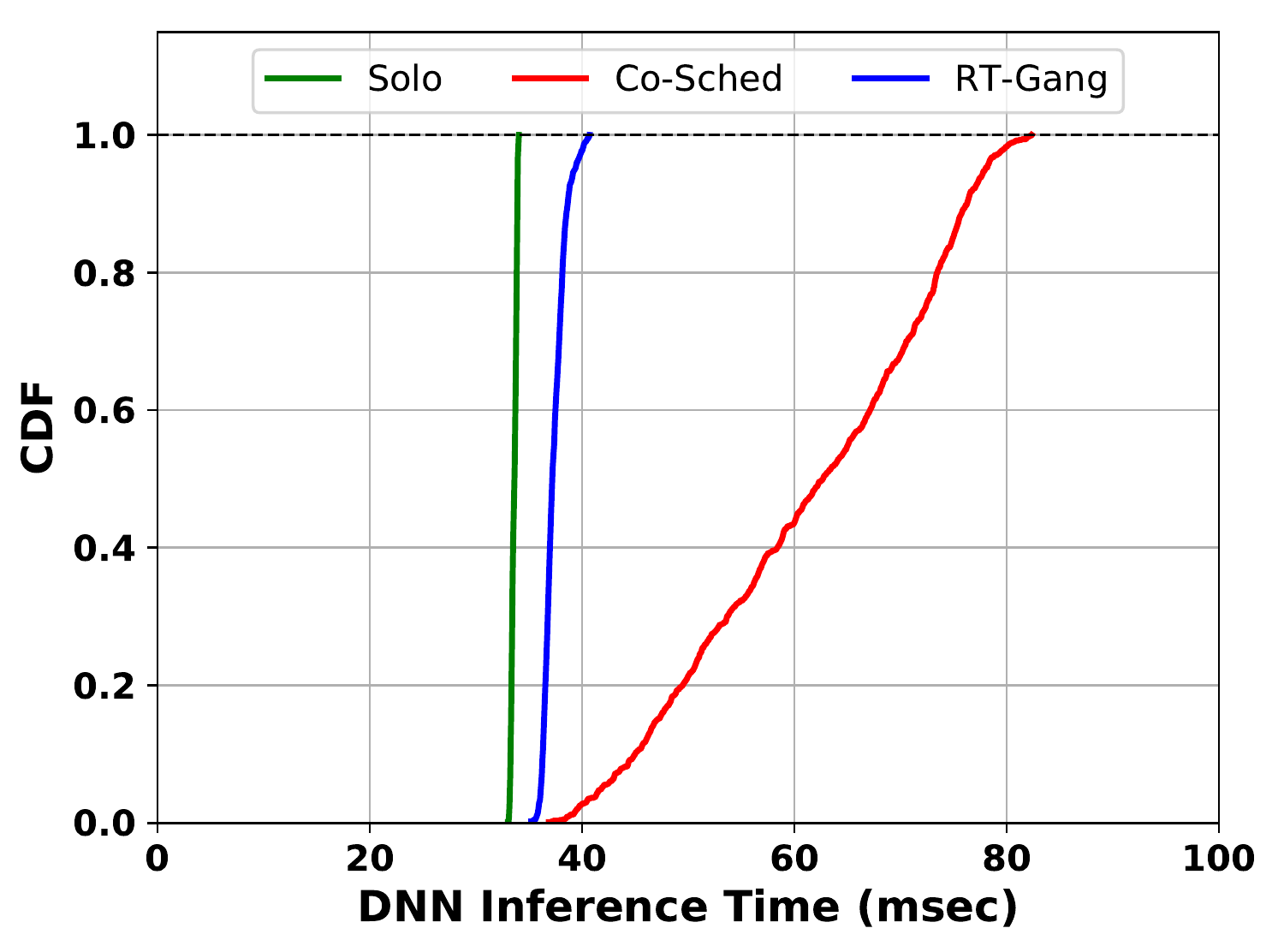}
	\label{fig:ev-pi_1n1c}
}
\subfigure[Pi3: DNN (3 Cores)]{
	\includegraphics[width=0.315\textwidth]{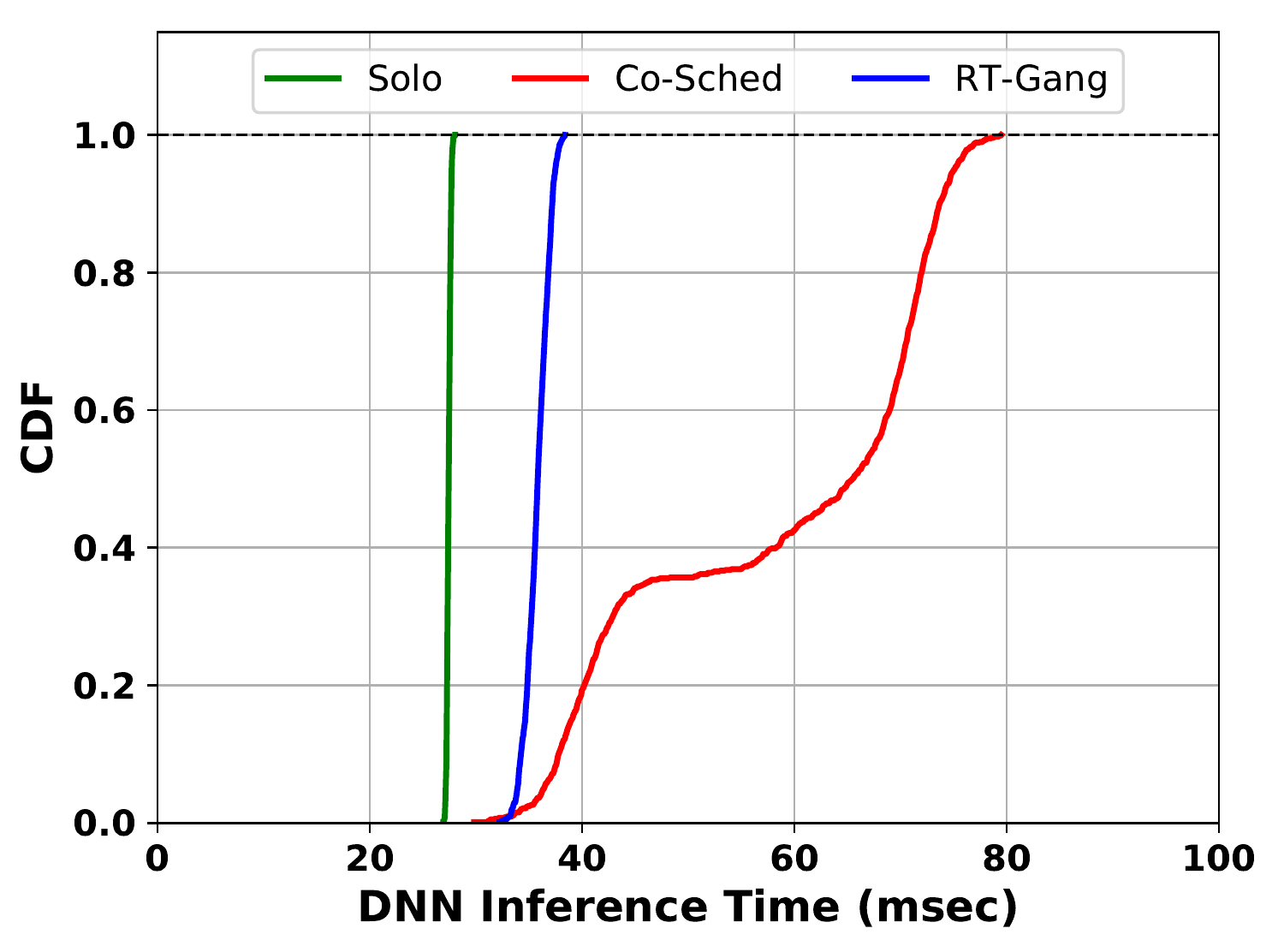}
	\label{fig:ev-pi_1n2c}
}
\subfigure[Pi3: DNN (4 Cores)]{
	\includegraphics[width=0.315\textwidth]{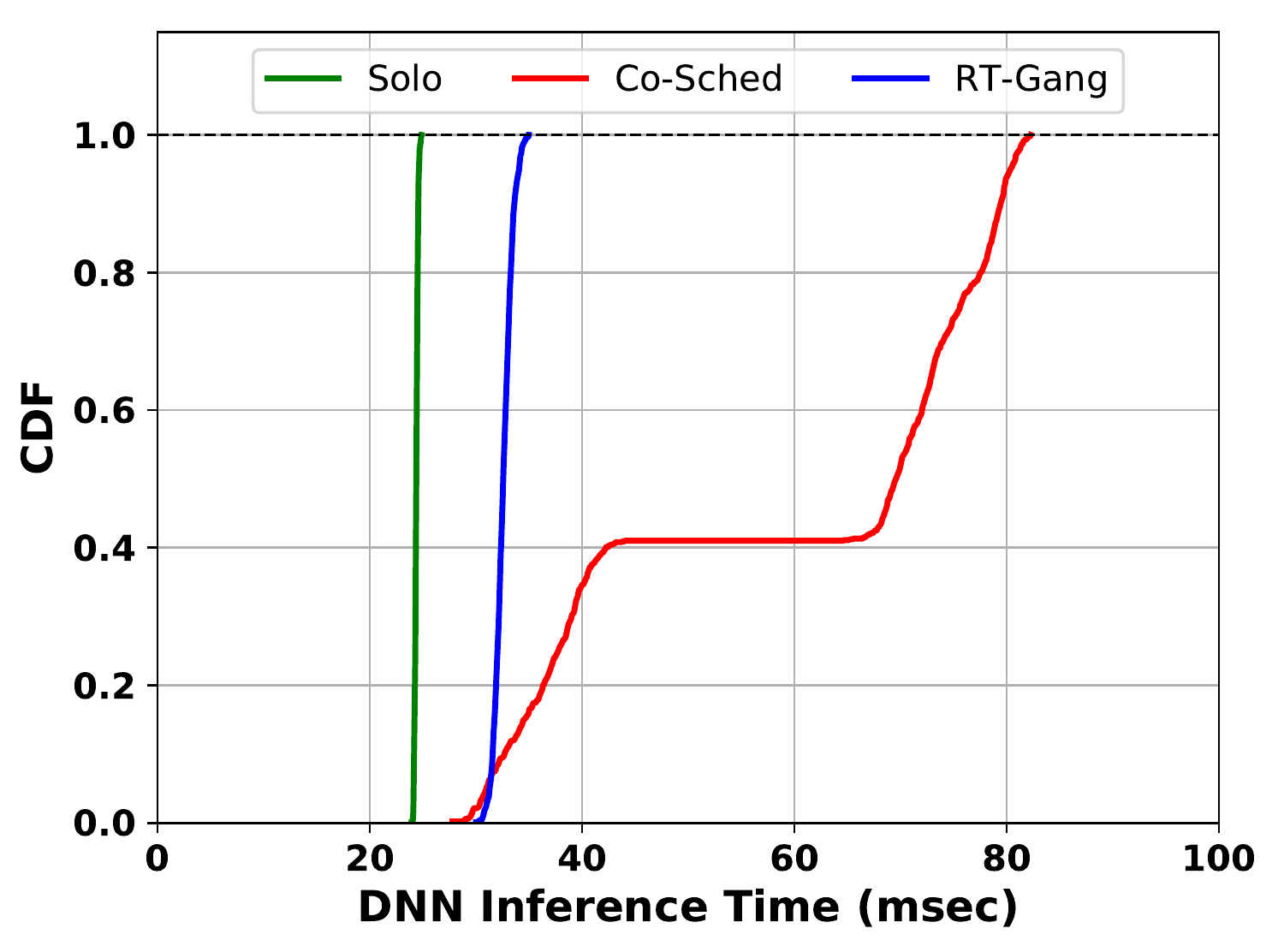}
	\label{fig:ev-pi_1n3c}
}
\caption{Performance of DNN inference loop under different scheduling schemes}
\label{fig:nn-eval}
\end{figure*}

\subsection{Synthetic Workload}\label{sec:eval_synthetic}
In this experiment, we show the benefits of RT-Gang using a synthetic 
taskset on Raspberry Pi3. The taskset is composed of two periodic
multi-threaded real-time tasks ($\tau_1$ and $\tau_2$) and two
single-threaded best-effort tasks ($\tau_{mem}^{BE}$ and
$\tau_{cpu}^{BE}$). Using this taskset, we explore task execution time
impacts to the real-time tasks and throughput impacts to the
best-effort tasks.

We use a modified version of the BwRead benchmark---which sequentially
accesses a 1-D array of a chosen size---from the
IsolBench~\cite{valsan2016taming} benchmark suite to construct the
taskset~\footnote{Our additions include support for multi-threaded and periodic
task invocation. The code can be found in the IsolBench
repository~\cite{isolbench}.}.
Concretely, $\tau_1$ creates two threads and is configured to use three
quarters of the last-level cache size (384KB out of 512KB L2 cache of
the Pi 3) as its working-set (which is shared between the two
threads). It is periodically released at a 20ms interval and
each job takes about 3.5ms in isolation. Similarly, $\tau_2$ is also a
dual-threaded periodic real-time task with the same working-set size
(384KB), but differs in its period (30 ms) and job execution times
(6.5ms in isolation). We set the
priority of $\tau_1$ higher than that of $\tau_2$, and schedule
 $\tau_1$ on Core 0,1 and $\tau_2$ on Core 2,3 (pinned using CPUSET
interface).

Note that $\tau_1$ and $\tau_2$ are scheduled by
SCHED\_FIFO real-time scheduler with and without RT-Gang
enabled~\footnote{RT-Gang can be enabled or disabled at runtime via: \texttt{/sys/kernel/debug/sched\_features}}.
On the other hand, $\tau_{mem}^{BE}$ and $\tau_{cpu}^{BE}$ are both
single-threaded best-effort tasks, which are scheduled by the CFS
scheduler~\cite{CFS}; they 
differ in that $\tau_{mem}^{BE}$'s working-set size is two times
bigger than the L2 cache size, while $\tau_{cpu}^{BE}$'s working-set
is smaller than the core's private L1 cache size. Thus,
$\tau_{mem}^{BE}$ may cause interference to co-scheduled real-time
tasks (if any) on different cores, due to contention in the shared L2 cache,
while $\tau_{cpu}^{BE}$ may not.
We collect the execution traces of the taskset without and with
RT-Gang for a duration of 10 seconds using the \texttt{trace-cmd} and 
KernelShark~\cite{kernelshark}.

Figure~\ref{fig:eval-gsched} shows the execution traces.  In inset (a), during
the first 20 msec duration, $\tau_1$ and $\tau_2$ were overlapped with each
other, which resulted in significant job execution time increase for both tasks
because their working-sets could not  fit into the shared L2 cache
simultaneously. During the next 20 msec duration, although $\tau_1$ and $\tau_2$
did not overlap, $\tau_1$ was overlapped with the two best-effort tasks,
$\tau_{mem}^{BE}$ and $\tau_{cpu}^{BE}$, which also resulted in a significant
increase in $\tau_1$'s job execution time (albeit to a lesser degree).

In inset (b), on the other hand, RT-Gang almost completely eliminates job
execution time variance. In the first 20 msec duration, $\tau_1$ is overlapped
with the two best-effort tasks. However, $\tau_{mem}^{BE}$ (the memory intensive
one) was throttled most of the time, which protected the execution time of
$\tau_1$. At around 40ms in the timeline, $\tau_1$ preempted $\tau_2$, the
real-time task. In place of $\tau_2$, two best-effort tasks were scheduled, of
which $\tau_{mem}^{BE}$ was again throttled as per $\tau_1$'s specified memory
bandwidth threshold setting.

Note that in
RT-Gang, the execution times of $\tau_1$ is
deterministic. Furthermore, $\tau_2$ is also highly predictable as we
only need to consider the preemption by the higher priority
$\tau_1$, according to the classic response time analysis
(RTA)\cite{Audsley93RTA}. Furthermore, because the two real-time tasks
do not experience significant execution time increases, there are more
``slacks'' left for the best-effort tasks---compared to without using
RT-Gang in inset (a)---which improves throughput of the best-effort
tasks.

\subsection{DNN Workload}
To establish the practicality of RT-Gang for real-world safety critical
applications, we used the DNN workload introduced in
Section~\ref{sec:motivation} and executed it under different
configurations on Raspberry Pi3 and NVIDIA Jetson TX2, with and without RT-Gang.
The Cortex-A53 cores in Raspberry Pi-3 are much less capable than the four
Cortex-A57 cores in Jetson TX2 platform. Moreover, the memory system in
Raspberry Pi3 offers significantly less bandwidth than the one in Jetson TX2.
Consequently, co-scheduled workloads in Raspberry Pi-3 are much more prone to
the negative effects of resource contention. By evaluating RT-Gang on these two
disparate platforms, we demonstrate its applicability to the range of embedded
devices available today.

{
\renewcommand{\arraystretch}{1.25}
\begin{table}[h]
	\centering
	\begin{tabularx}{0.8\linewidth}{l|l|l|l}
		\toprule
		Task 			& WCET ($C$ ms)	&   Period
                ($P$ ms)& \# Threads \\
        \hline
        \multicolumn{4}{c}{Common}\\
        \hline
		$\tau_{cutcp}^{BE}$		& $\infty$		    & N/A		    & 4\\
		$\tau_{lbm}^{BE}$		& $\infty$		    & N/A		    & 4\\
        \hline
        \multicolumn{4}{c}{Jetson TX2}\\
        \hline
		$\tau_{bww}^{RT}$		& 40.0 			    & 100.0		    & 4\\
		$\tau_{dnn(2)}^{RT}$	& 10.7			    & 24.0	        & 2\\
		$\tau_{dnn(3)}^{RT}$	& 8.8 			    & 19.0	        & 3\\
		$\tau_{dnn(4)}^{RT}$	& 7.6 			    & 17.0	        & 4\\
        \hline
        \multicolumn{4}{c}{Raspberry Pi 3}\\
        \hline
		$\tau_{bww}^{RT}$		& 47.0 			    & 100.0		    & 4\\
		$\tau_{dnn(2)}^{RT}$	& 34.0			    & 78.0	        & 2\\
		$\tau_{dnn(3)}^{RT}$	& 27.90 		    & 65.0	        & 3\\
		$\tau_{dnn(4)}^{RT}$	& 24.81 		    & 56.0	        & 4\\
		\bottomrule
	\end{tabularx}
	\caption{Taskset parameters for the DNN experiment.}
	\label{tbl:dnn_eval}
\end{table}
}

The taskset parameters for this experiment are shown in
Table~\ref{tbl:dnn_eval}. First, we use a multi-threaded DNN application as the high
priority periodic real-time task ($\tau_{dnn(c)}^{RT}$ where $c$
denotes the number of cores used). The period of DNN inference
operation is selected to keep the per-core utilization of DNN threads
close to 45\%.
Second, as a lower priority real-time task,
we use a periodic multi-threaded BwWrite benchmark ($\tau_{bww}^{RT}$). The
working set size of BwWrite was chosen to be twice the size of LLC in
each platform so as to stress memory subsystem (due to cache misses).
The compute time ($C$) of BwWrite is selected to keep its
per-core utilization less than 50\% in all experiments.
Lastly, we use two benchmarks from Parboil suite~\cite{parboil}, \texttt{lbm}
($\tau_{lbm}^{BE}$) and \texttt{cutcp},  ($\tau_{cutcp}^{BE}$)) as
best-effort tasks, which represent memory and CPU intensive parallel OpenMP
applications respectively. 

We vary the thread count (= \# of assigned CPU cores)
of the DNN task while keeping the thread count of $\tau_{bww}^{RT}$
and the best-effort tasks ($\tau_{lbm}^{BE}$ and $\tau_{cutcp}^{BE}$)
fixed at four.
For the experiment, we first measure the performance of
$\tau_{dnn(c)}^{RT}$ in isolation, then co-scheduled with the rest of
the taskset on baseline Linux, and finally using RT-Gang.

Figure~\ref{fig:nn-eval} shows the cumulative distribution function
(CDF) of the per-frame DNN inference time in each configuration
(\emph{Solo}: alone in isolation, \emph{Co-Sched}: co-scheduled under
baseline Linux, \emph{RT-Gang}: co-scheduled under RT-Gang enabled Linux).
Note first that execution times of the DNN task vary 
significantly under the co-scheduling scheme (\emph{Co-Sched}).
On Raspberry Pi3, the WCET across all configurations is more than
2X of its solo WCET.
On TX2, the co-scheduling graphs again show deteriorated performance, albeit
to a lesser extent than Raspberry Pi3.

Under RT-Gang, on the other hand, the execution times of the
DNN workload are highly deterministic and match closely with its solo
execution times in all tested configurations on both platforms.
On Raspberry Pi3, however, although DNN's performance is
deterministic (i.e., low variance), noticeable performance constant
increase is observed under RT-Gang when compared to the solo execution
of the DNN task. We believe that this is caused by Cache Related Preemption Delay
(CRPD)~\cite{lee1998analysis},
as the memory system of Raspberry Pi3 is significantly less powerful
than that of Jetson TX2.

Note that taking CRPD into analysis is well known in the context of
single-core processors, but its applications have been difficult in
multicore due to cache contention among the co-scheduled tasks on
different cores, as shown in the CDF plots of \emph{Co-Sched}. The
determinism that RT-Gang brings thus would make CRPD analysis valid on
multicore processors again, enabling effective timing analysis.

\subsection{Overhead}
There are two main sources of overhead in our implementation of RT-Gang: 1)
The serialization overhead associated with the critical section of our
gang scheduling algorithm in selecting next real-time task. 2) The overhead
involved in sending cross-core interrupts (IPIs) and acquiring ready-queue
locks for gang preemption.

The serialization overhead of RT-Gang is only incurred during the selection of
real-time tasks due to the use of a spinlock. However, the length of the
critical section of RT-Gang is small---comparable to existing spinlocks used in
the various parts of the Linux kernel scheduler.  On the other hand, the
overhead associated with gang-preemption due to the IPIs can pose a scalability
problem as the number of necessary IPIs can be as many as all the rest of the
cores.

In order to estimate both these overheads, we conducted an experiment
on the NVIDIA Jetson TX2 platform in which a high priority real-time
gang preempts a multi-threaded low priority real-time gang, a fixed
number of times ($100000$), with and without RT-Gang. We also varied
the number of threads of the low priority gang to see the effect of
gang size on the gang preemption overhead. The result from this
experiment is shown in Table~\ref{tbl:ovhead}. 

{
\renewcommand{\arraystretch}{1.25}
\begin{table}[h]
	\centering
	\begin{tabularx}{0.9\linewidth}{c|c}
		\toprule
		Scenario			& Context Switch Cost (usec)\\
		\midrule
		1-Thread-Lowprio (Linux)	& 6.81 \\
		1-Thread-Lowprio (RT-Gang)	& 7.19 \\
		2-Thread-Lowprio (RT-Gang)	& 7.37 \\
		3-Thread-Lowprio (RT-Gang)	& 7.55 \\
		4-Thread-Lowprio (RT-Gang)	& 7.72 \\
		\bottomrule
	\end{tabularx}
	\caption{RT-Gang Overhead in Linux}
	\label{tbl:ovhead}
\end{table}
}

As can been seen from the table, RT-Gang adds very small overhead to the overall
cost of a context-switch under Linux; considering the fact that for a
well-designed system, a context-switch is not supposed to happen too
frequently. The findings from this experiment also match the results seen
during evaluation with DNN workloads; in which, we saw that the performance of
these workloads remain completely unaffected under RT-Gang.

 \section{Discussion}\label{sec:discussion}
In this section, we briefly discuss potential use-cases of RT-Gang.
We believe that our simple design and practical implementation leveraging
existing real-time scheduler in Linux offer broader practical
use-cases in many application domains that concern timing
predictability and need to run parallelized multi-thread applications.
Time critical parallel real-time applications in automotive and
aviation domains (e.g., perception and control applications in a
self-driving car) are our main target use-cases. Also, barrier based
scientific applications in high-performance computing (HPC) can potentially
benefit from using RT-Gang as they are sensitive to thread imbalance,
and thus motivated original gang scheduling
research~\cite{feitelson1992gang} in the first 
place. Although we mainly target embedded multicore
processors (typically having 4-8 cores) in this work, we recently were
able to apply our RT-Gang kernel patch on a 12 hardware thread (6
core) x86 PC, successfully performing gang scheduling across the 12
hardware threads. Effectiveness and scalability of RT-Gang in other
application domains, such as HPC, is left as future work.

 \section{Related Work}\label{sec:related}

{\bf Parallel Real-Time Scheduling.}
Parallel real-time tasks are generally modeled using one of following
three models: Fork-join model~\cite{lakshmanan2010scheduling,saifullah2013rtj_parallel, nelissen2012techniques,chwa2013global}, dag model~\cite{baruah2012rtss_generalized,saifullah2014tpds_parallel} and gang task model~\cite{goossens2010rtns_gangftp,kato2009rtss_gangedf,dong2017rtss_analysis}. In the
fork-join model, a task alternates parallel (fork) and sequential
(join) phases over time. In the dag model, a task is represented as a
directed acyclic graph with a set of associated precedence
constraints, which allows more flexible scheduling as long as the
constraints are satisfied. Lastly, in the gang model, a task is simply
represented with an execution time $e$ and a number of cores $k$ it
needs to run. While most simple among the three, it matches well with
real-world parallel applications where users or scheduler selects the
number of threads an application might use at the start time, which
is also the case in DNN workload we used in Section~\ref{sec:motivation}.

Our one-gang-at-a-time scheduling policy is essentially a restrictive
form of gang scheduling~\cite{feitelson1992gang}. In traditional gang
scheduling, all threads of an application are scheduled simultaneously, and more
than one application is allowed to be scheduled as long as there are
available cores. In contrast, our approach is restrictive in the sense that such
co-scheduling is not allowed. Gang scheduling was originally studied
in high-performance computing as a way to maximize performance of
parallel applications~\cite{feitelson1992gang} or virtual machine
scheduling~\cite{gangcfs}. More recently, gang scheduling was
investigated as a way to improve security
\cite{sprabery2018scheduling}, by preventing simultaneously co-scheduling 
different security domains, similar to our one-gang-at-a-time
policy, but it was implemented in Linux's CFS scheduler and thus does
not support real-time task scheduling.  

In the real-time systems community,
fixed-priority and dynamic priority real-time versions of gang
scheduling policies, namely Gang FTP and Gang EDF, respectively, are
studied and
analyzed~\cite{goossens2010rtns_gangftp,kato2009rtss_gangedf,dong2017rtss_analysis}.
However, these prior real-time gang scheduling policies also allow co-scheduling of
multiple gangs as long as there exist available cores because they
mainly focus on CPU utilization, without considering the negative WCET
impact of co-scheduling on shared memory multicore platforms.
On the other hand, the Isolation Scheduling
model~\cite{huang2015isolation} and the integrated modular avionic
(IMA) scheduler design in~\cite{melani2017scheduling} 
consider shared resource interference and limit co-scheduling to the
tasks of the same criticality (in ~\cite{huang2015isolation}) or 
those in the same IMA partition (in~\cite{melani2017scheduling}).
However, they do not specifically target parallel real-time tasks and do not
allow co-scheduling of best-effort tasks.
Also, to our knowledge, all aforementioned real-time
scheduling policies were not implemented in actual OSes.
Our work is, to the best of our knowledge, the first real-time gang
scheduler implementation in Linux, which implements a fixed-priority
real-time gang scheduling, enforcing the one-gang-at-a-time policy.
Furthermore, we enable safely co-scheduling best-effort tasks, by
integrating an OS-level throttling mechanism~\cite{yun2017bwlock},
to improve system utilization when there are best-effort tasks---a
common situation in practice.

{\bf Real-Time Scheduling in Linux.}
A number of efforts in the real-time systems community have been aimed
at improving real-time scheduling in Linux. For example,
$LITMUS^{RT}$~\cite{calandrino2006litmus} developed a set of global,
partitioned, and semi-partitioned real-time schedulers, outside of the existing
Linux's real-time scheduler; ChronOS Linux developed global and local scheduling
layers built on top of Linux's O(1) scheduler~\cite{dellinger2011chronos}.
In contrast, RT-Gang is
developed as an extension (a feature) of Linux's default real-time
scheduler that enforces a simple invariant---one 
real-time gang across all cores---otherwise respecting standard fixed-priority
real-time scheduling, and does not maintain its own scheduling queues, unlike
the aforementioned two prior implementations. We believe our design is
simpler and easier to maintain in the face of rapid change in Linux kernel.

{\bf OS-level Shared Resource Partitioning.}
Many researchers have attempted to make COTS
multicore platforms to be more predictable with OS-level techniques.
A majority of prior works focused on \emph{partitioning} of shared
resources among the tasks and cores to improve predictability.
Page coloring has long been studied to
partition shared cache~\cite{liedtke97ospart,lin2008gaining,zhang2009towards,soares2008reducing,ding2011srm,ward2013ecrts,kim2013coordinated,ye2014coloris},
DRAM banks~\cite{yun2014rtas,liu2012software,suzuki2013coordinated},
and TLB~\cite{panchamukhi2015providing}.
Some COTS processors~\cite{intelcat,mancuso2013rtas} support cache-way
partitioning~\cite{suh2002new}. Mancuso et al.~\cite{mancuso2013rtas}
and Kim et al.~\cite{kim2017attacking}, used both coloring and
cache way partitioning for fine-grained cache partitioning.
While these shared resource partitioning 
techniques can reduce space conflicts of some shared resources, hence
beneficial for predictability, but they are not enough to guarantee
strong time predictability on COTS multicore platforms because there are too many
hardware resources (e.g., cache MSHRs, DRAM controller buffers, etc.)
that have profound impact on task timing~\cite{valsan2016taming,michael2019dos}, but
are unpartitionable and out of control of software.

 \section{Conclusion}\label{sec:conclusion}
We presented RT-Gang: a novel real-time gang scheduling
framework for predictable and efficient parallel real-time scheduling
on multicore.

RT-Gang implements a novel gang scheduling policy that eliminates
inter-task interference by enforcing an invariant that 
only one (parallel) real-time task (gang) can be scheduled at any given
time. This enables tighter task WCET estimation and
simpler schedulability analysis.
RT-Gang also provides additional mechanisms, namely virtual gang and
best-effort task throttling, which can help maximize system utilization 
while providing strong time predictability to real-time tasks.

We implemented RT-Gang in Linux and evaluated it on two
embedded multicore platforms. The evaluation results
show the predictability and efficiency benefits of RT-Gang.
In future, we plan to extend our RT-Gang approach to heterogeneous
multicore platforms.

\balance
 \section*{Acknowledgements} \label{acknowledge}
This research is supported by NSF CNS 1718880, CNS 1815959, and NSA Science of Security initiative contract \#H98230-18-D-0009.
 \bibliographystyle{abbrv}
\bibliography{research}

\begin{thebibliography}{10}

\bibitem{isolbench}
{IsolBench code repository}.
\newblock \url{https://github.com/CSL-KU/IsolBench}.

\bibitem{rt-gang}
{RT-Gang code repository}.
\newblock \url{https://github.com/CSL-KU/RT-Gang}.

\bibitem{ali2018protecting}
W.~Ali and H.~Yun.
\newblock {Protecting Real-Time GPU Kernels on Integrated CPU-GPU SoC
  Platforms}.
\newblock In {\em Euromicro Conference on Real-Time Systems (ECRTS)}, 2018.

\bibitem{Audsley93RTA}
N.~Audsley, A.~Burns, M.~Richardson, K.~Tindell, and A.~Wellings.
\newblock Applying new scheduling theory to static priority preemptive
  scheduling.
\newblock {\em Software Engineering Journal}, 8(5):284--292, 1993.

\bibitem{axer2014building}
P.~Axer, R.~Ernst, H.~Falk, A.~Girault, D.~Grund, N.~Guan, B.~Jonsson,
  P.~Marwedel, J.~Reineke, C.~Rochange, M.~Sebastian, R.~V. Hanxleden,
  R.~Wilhelm, and W.~Yi.
\newblock Building timing predictable embedded systems.
\newblock {\em ACM Transactions on Embedded Computing Systems (TECS)},
  13(4):82:1--82:37, 2014.

\bibitem{baruah2012rtss_generalized}
S.~Baruah, V.~Bonifaci, A.~Marchetti-Spaccamela, L.~Stougie, and A.~Wiese.
\newblock A generalized parallel task model for recurrent real-time processes.
\newblock In {\em Real-Time Systems Symposium (RTSS)}, pages 63--72. IEEE,
  2012.

\bibitem{michael2018deep}
M.~G. Bechtel, E.~McEllhiney, and H.~Yun.
\newblock {DeepPicar: A Low-cost Deep Neural Network-based Autonomous Car}.
\newblock In {\em Embedded and Real-Time Computing Systems and Applications
  (RTCSA)}, 2018.

\bibitem{michael2019dos}
M.~G. Bechtel and H.~Yun.
\newblock Denial-of-service attacks on shared cache in multicore: Analysis and
  prevention.
\newblock In {\em Real-Time and Embedded Technology and Applications Symposium
  (RTAS)}, 2019.

\bibitem{Bojarski2016}
M.~Bojarski, D.~D. Testa, D.~Dworakowski, B.~Firner, B.~Flepp, P.~Goyal, L.~D.
  Jackel, M.~Monfort, U.~Muller, J.~Zhang, X.~Zhang, J.~Zhao, and K.~Zieba.
\newblock {End to End Learning for Self-Driving Cars}.
\newblock {\em CoRR}, abs/1604.07316, 2016.

\bibitem{calandrino2006litmus}
J.~M. Calandrino, H.~Leontyev, A.~Block, U.~C. Devi, and J.~H. Anderson.
\newblock {LITMUS\^{} RT: A Testbed for Empirically Comparing Real-Time
  Multiprocessor Schedulers}.
\newblock In {\em Real-Time Systems Symposium, 2006. RTSS'06. 27th IEEE
  International}, pages 111--126. IEEE, 2006.

\bibitem{faa2014certification}
{Certification Authorities Software Team}.
\newblock {CAST-32: Multi-core Processors (Rev 0)}.
\newblock Technical report, Federal Aviation Administration (FAA), May 2014.

\bibitem{faa2016certification}
{Certification Authorities Software Team}.
\newblock {CAST-32A: Multi-core Processors}.
\newblock Technical report, Federal Aviation Administration (FAA), November
  2016.

\bibitem{chwa2013global}
H.~S. Chwa, J.~Lee, K.-M. Phan, A.~Easwaran, and I.~Shin.
\newblock {Global edf schedulability analysis for synchronous parallel tasks on
  multicore platforms}.
\newblock In {\em Euromicro Conference on Real-Time Systems (ECRTS)}, pages
  25--34. IEEE, 2013.

\bibitem{gangcfs}
N.~A. Dadhania.
\newblock {Gang scheduling in CFS}.
\newblock \url{https://lwn.net/Articles/472797/}.

\bibitem{dellinger2011chronos}
M.~Dellinger, P.~Garyali, and B.~Ravindran.
\newblock {ChronOS Linux: a best-effort real-time multiprocessor Linux kernel}.
\newblock In {\em Proceedings of the 48th Design Automation Conference}, pages
  474--479. ACM, 2011.

\bibitem{ding2011srm}
X.~Ding, K.~Wang, and X.~Zhang.
\newblock Srm-buffer: An os buffer management technique to prevent last level
  cache from thrashing in multicores.
\newblock In {\em Proceedings of the Sixth Conference on Computer Systems},
  EuroSys, pages 243--256, 2011.

\bibitem{dong2017rtss_analysis}
Z.~Dong and C.~Liu.
\newblock {Analysis Techniques for Supporting Hard Real-Time Sporadic Gang Task
  Systems}.
\newblock In {\em Real-Time Systems Symposium (RTSS)}, pages 128--138, 2017.

\bibitem{feitelson1992gang}
D.~G. Feitelson and L.~Rudolph.
\newblock Gang scheduling performance benefits for fine-grain synchronization.
\newblock {\em Journal of Parallel and distributed Computing}, 16(4):306--318,
  1992.

\bibitem{goossens2010rtns_gangftp}
J.~Goossens and V.~Berten.
\newblock {Gang FTP scheduling of periodic and parallel rigid real-time tasks}.
\newblock In {\em International Conference on Real-Time Networks and Systems
  (RTNS)}, pages 189--196, 2010.

\bibitem{bosch2019challenge}
A.~Hamann.
\newblock Industrial challenges: Moving from classical to high performance
  real-time systems.
\newblock In {\em International Workshop on Analysis Tools and Methodologies
  for Embedded and Real-time Systems (WATERS)}, July 2018.

\bibitem{huang2015isolation}
P.~Huang, G.~Giannopoulou, R.~Ahmed, D.~B. Bartolini, and L.~Thiele.
\newblock An isolation scheduling model for multicores.
\newblock In {\em Real-Time Systems Symposium (RTSS)}, pages 141--152. IEEE,
  2015.

\bibitem{intelcat}
Intel.
\newblock Improving real-time performance by utilizing cache allocation
  technology.
\newblock
  \url{https://software.intel.com/en-us/articles/introduction-to-cache-allocation-technology}.

\bibitem{kato2009rtss_gangedf}
S.~Kato and Y.~Ishikawa.
\newblock {Gang EDF scheduling of parallel task systems}.
\newblock In {\em Real-Time Systems Symposium (RTSS)}, pages 459--468. IEEE,
  2009.

\bibitem{kim2013coordinated}
H.~Kim, A.~Kandhalu, and R.~Rajkumar.
\newblock A coordinated approach for practical os-level cache management in
  multi-core real-time systems.
\newblock In {\em Euromicro Conference on Real-Time Systems (ECRTS)}, pages
  80--89, 2013.

\bibitem{kim2017attacking}
N.~Kim, B.~C. Ward, M.~Chisholm, J.~H. Anderson, and F.~D. Smith.
\newblock Attacking the one-out-of-m multicore problem by combining hardware
  management with mixed-criticality provisioning.
\newblock {\em Real-Time Systems}, 53(5):709--759, 2017.

\bibitem{lakshmanan2010scheduling}
K.~Lakshmanan, S.~Kato, and R.~Rajkumar.
\newblock Scheduling parallel real-time tasks on multi-core processors.
\newblock In {\em Real-Time Systems Symposium (RTSS)}, pages 259--268. IEEE,
  2010.

\bibitem{lee1998analysis}
C.-G. Lee, H.~Hahn, Y.-M. Seo, S.~L. Min, R.~Ha, S.~Hong, C.~Y. Park, M.~Lee,
  and C.~S. Kim.
\newblock Analysis of cache-related preemption delay in fixed-priority
  preemptive scheduling.
\newblock {\em IEEE transactions on computers}, 47(6):700--713, 1998.

\bibitem{lehoczky1989rate}
J.~Lehoczky, L.~Sha, and Y.~Ding.
\newblock The rate monotonic scheduling algorithm: exact characterization and
  average case behavior.
\newblock In {\em Real-Time Systems Symposium (RTSS)}, pages 166--171, 1989.

\bibitem{robert2015keynote}
R.~Leibinger.
\newblock {Software Architectures for Advanced Driver Assistance Systems
  (ADAS)}.
\newblock In {\em International Workshop on Operating Systems Platforms for
  Embedded Real-Time Applications (OSPERT)}, 2015.

\bibitem{liedtke97ospart}
J.~Liedtke, H.~Hartig, and M.~Hohmuth.
\newblock Os-controlled cache predictability for real-time systems.
\newblock In {\em IEEE Real-Time and Embedded Technology and Applications
  Symposium (RTAS)}, pages 213--224, 1997.

\bibitem{lin2008gaining}
J.~Lin, Q.~Lu, X.~Ding, Z.~Zhang, X.~Zhang, and P.~Sadayappan.
\newblock Gaining insights into multicore cache partitioning: Bridging the gap
  between simulation and real systems.
\newblock In {\em IEEE International Symposium on High Performance Computer
  Architecture (HPCA)}, pages 367--378, 2008.

\bibitem{liu2012software}
L.~Liu, Z.~Cui, M.~Xing, Y.~Bao, M.~Chen, and C.~Wu.
\newblock A software memory partition approach for eliminating bank-level
  interference in multicore systems.
\newblock In {\em International Conference on Parallel Architectures and
  Compilation Techniques (PACT)}, pages 367--375, 2012.

\bibitem{mancuso2013rtas}
R.~Mancuso, R.~Dudko, E.~Betti, M.~Cesati, M.~Caccamo, and R.~Pellizzoni.
\newblock Real-time cache management framework for multi-core architectures.
\newblock In {\em IEEE Real-Time and Embedded Technology and Applications
  Symposium (RTAS)}, pages 45--54, 2013.

\bibitem{melani2017scheduling}
A.~Melani, R.~Mancuso, M.~Caccamo, G.~Buttazzo, J.~Freitag, and S.~Uhrig.
\newblock A scheduling framework for handling integrated modular avionic
  systems on multicore platforms.
\newblock In {\em Embedded and Real-Time Computing Systems and Applications
  (RTCSA)}, pages 1--10. IEEE, 2017.

\bibitem{CFS}
I.~Molnar.
\newblock Cfs scheduler.
\newblock
  \url{https://www.kernel.org/doc/Documentation/scheduler/sched-design-CFS.txt}.

\bibitem{nelissen2012techniques}
G.~Nelissen, V.~Berten, J.~Goossens, and D.~Milojevic.
\newblock Techniques optimizing the number of processors to schedule
  multi-threaded tasks.
\newblock In {\em Euromicro Conference on Real-Time Systems (ECRTS)}, pages
  321--330. IEEE, 2012.

\bibitem{nvidia2017bb8}
NVIDIA.
\newblock {NVIDIA BB8 Self-Driving Car}.
\newblock \url{https://blogs.nvidia.com/blog/2017/01/04/bb8-ces/}, 2017.

\bibitem{panchamukhi2015providing}
S.~A. Panchamukhi and F.~Mueller.
\newblock Providing task isolation via tlb coloring.
\newblock In {\em IEEE Real-Time and Embedded Technology and Applications
  Symposium (RTAS)}, pages 3--13, 2015.

\bibitem{pellizzoni2016memory}
R.~Pellizzoni and H.~Yun.
\newblock Memory servers for multicore systems.
\newblock In {\em IEEE Real-Time and Embedded Technology and Applications
  Symposium (RTAS)}, pages 1--12, 2016.

\bibitem{kernelshark}
S.~Rostedt.
\newblock Kernelshark.
\newblock \url{http://rostedt.homelinux.com/kernelshark/}.

\bibitem{saifullah2014tpds_parallel}
A.~Saifullah, D.~Ferry, J.~Li, K.~Agrawal, C.~Lu, and C.~D. Gill.
\newblock {Parallel real-time scheduling of DAGs}.
\newblock {\em Parallel and Distributed Systems, IEEE Transactions on},
  25(12):3242--3252, 2014.

\bibitem{saifullah2013rtj_parallel}
A.~Saifullah, J.~Li, K.~Agrawal, C.~Lu, and C.~Gill.
\newblock {Multi-core real-time scheduling for generalized parallel task
  models}.
\newblock {\em Real-Time Systems}, 49(4):404--435, 2013.

\bibitem{sha2004real}
L.~Sha, T.~Abdelzaher, K.-E. AArzen, A.~Cervin, T.~Baker, A.~Burns,
  G.~Buttazzo, M.~Caccamo, J.~Lehoczky, and A.~K. Mok.
\newblock Real time scheduling theory: A historical perspective.
\newblock {\em Real-Time Systems}, 28(2-3):101--155, 2004.

\bibitem{skalistis2018safe}
S.~Skalistis, F.~Angiolini, G.~De~Micheli, and A.~Simalatsar.
\newblock Safe and efficient deployment of data-parallelizable applications on
  many-core platforms: Theory and practice.
\newblock {\em IEEE Design \& Test}, 35(4):7--15, 2018.

\bibitem{soares2008reducing}
L.~Soares, D.~Tam, and M.~Stumm.
\newblock Reducing the harmful effects of last-level cache polluters with an
  os-level, software-only pollute buffer.
\newblock In {\em IEEE/ACM International Symposium on Microarchitecture
  (MICRO)}, pages 258--269, 2008.

\bibitem{sprabery2018scheduling}
R.~Sprabery, K.~Evchenko, A.~Raj, R.~B. Bobba, S.~Mohan, and R.~Campbell.
\newblock Scheduling, isolation, and cache allocation: A side-channel defense.
\newblock In {\em Cloud Engineering (IC2E), International Conference on}, pages
  34--40. IEEE, 2018.

\bibitem{sprunt1989aperiodic}
B.~Sprunt.
\newblock {\em Aperiodic Task Scheduling for Real-time Systems}.
\newblock PhD thesis, 1990.
\newblock AAI9107570.

\bibitem{parboil}
J.~A. Stratton, C.~Rodrigues, I.-J. Sung, N.~Obeid, L.-W. Chang, N.~Anssari,
  G.~D. Liu, and W.~mei W.~Hwu.
\newblock Parboil: A revised benchmark suite for scientific and commercial
  throughput computing.
\newblock Technical report, University of Illinois at Urbana-Champaign, 2012.

\bibitem{suh2002new}
G.~E. Suh, S.~Devadas, and L.~Rudolph.
\newblock A new memory monitoring scheme for memory-aware scheduling and
  partitioning.
\newblock In {\em International Symposium on High Performance Computer
  Architecture}, pages 117--128, 2002.

\bibitem{suzuki2013coordinated}
N.~Suzuki, H.~Kim, D.~d.~Niz, B.~Andersson, L.~Wrage, M.~Klein, and
  R.~Rajkumar.
\newblock Coordinated bank and cache coloring for temporal protection of memory
  accesses.
\newblock In {\em IEEE International Conference on Computational Science and
  Engineering (CSE)}, pages 685--692, 2013.

\bibitem{valsan2016taming}
P.~K. Valsan, H.~Yun, and F.~Farshchi.
\newblock Taming non-blocking caches to improve isolation in multicore
  real-time systems.
\newblock In {\em Real-Time and Embedded Technology and Applications Symposium
  (RTAS)}, 2016.

\bibitem{valsan2017addressing}
P.~K. Valsan, H.~Yun, and F.~Farshchi.
\newblock Addressing isolation challenges of non-blocking caches for multicore
  real-time systems.
\newblock {\em Real-Time Systems}, 53(5):673--708, 2017.

\bibitem{ward2013ecrts}
B.~C. Ward, J.~L. Herman, C.~J. Kenna, and J.~H. Anderson.
\newblock Making shared caches more predictable on multicore platforms.
\newblock In {\em Euromicro Conference on Real-Time Systems (ECRTS)}, pages
  157--167, 2013.

\bibitem{ye2014coloris}
Y.~Ye, R.~West, Z.~Cheng, and Y.~Li.
\newblock Coloris: A dynamic cache partitioning system using page coloring.
\newblock In {\em International Conference on Parallel Architecture and
  Compilation Techniques (PACT)}, pages 381--392, 2014.

\bibitem{yun2017bwlock}
H.~Yun, W.~Ali, S.~Gondi, and S.~Biswas.
\newblock {BWLOCK: A Dynamic Memory Access Control Framework for Soft Real-Time
  Applications on Multicore Platforms}.
\newblock {\em IEEE Transactions on Computers (TC)}, PP(99):1--1, 2016.

\bibitem{yun2014rtas}
H.~Yun, R.~Mancuso, Z.~Wu, and R.~Pellizzoni.
\newblock {PALLOC: DRAM bank-aware memory allocator for performance isolation
  on multicore platforms}.
\newblock In {\em IEEE Real-Time and Embedded Technology and Applications
  Symposium (RTAS)}, pages 155--166, 2014.

\bibitem{yun2015ospert}
H.~Yun and P.~K. Valsan.
\newblock Evaluating the isolation effect of cache partitioning on cots
  multicore platforms.
\newblock In {\em Workshop on Operating Systems Platforms for Embedded
  Real-Time Applications (OSPERT)}, 2015.

\bibitem{zhang2009towards}
X.~Zhang, S.~Dwarkadas, and K.~Shen.
\newblock Towards practical page coloring-based multicore cache management.
\newblock In {\em Proceedings of the 4th ACM European Conference on Computer
  Systems}, EuroSys '09, pages 89--102, 2009.

\end{thebibliography}

\end{document}